# Revisiting Frequency Moment Estimation in Random Order Streams

Vladimir Braverman[*],  
Johns Hopkins University

Emanuele Viola[†],  
Northeastern University

David Woodruff[‡],  
Carnegie Mellon University

Lin F. Yang[§]  
Princeton University

March 6, 2018

## Abstract

We revisit one of the classic problems in the data stream literature, namely, that of estimating the frequency moments $F_p$ for $0 < p < 2$ of an underlying $n$-dimensional vector presented as a sequence of additive updates in a stream. It is well-known that using $p$-stable distributions one can approximate any of these moments up to a multiplicative $(1 + \epsilon)$-factor using $O(\epsilon^{-2} \log n)$ bits of space, and this space bound is optimal up to a constant factor in the turnstile streaming model. We show that surprisingly, if one instead considers the popular random-order model of insertion-only streams, in which the updates to the underlying vector arrive in a random order, then one can beat this space bound and achieve $\widetilde{O}(\epsilon^{-2} + \log n)$ bits of space, where the $\widetilde{O}$ hides $\mathrm{poly}(\log(1/\epsilon) + \log \log n)$ factors. If $\epsilon^{-2} \approx \log n$, this represents a roughly quadratic improvement in the space achievable in turnstile streams. Our algorithm is in fact deterministic, and we show our space bound is optimal up to $\mathrm{poly}(\log(1/\epsilon) + \log \log n)$ factors for deterministic algorithms in the random order model. We also obtain a similar improvement in space for $p = 2$ whenever $F_2 \gtrsim \log n \cdot F_1$.

## 1 Introduction

Analyzing massive datasets has become an increasingly challenging problem. Data sets, such as sensor networks, stock data, web/network traffic, and database transactions, are collected at a tremendous pace. Traditional algorithms that store an entire dataset in memory are impractical. The *streaming model* has emerged as an important model for coping with massive datasets. Streaming algorithms are typically randomized and approximate, making a single pass over the data and using only a small amount of memory.

[*]vova@cs.jhu.edu  
[†]viola@ccs.neu.edu  
[‡]dwoodruf@cs.cmu.edu  
[§]lin.yang@princeton.edu. Work done while the author was at Johns Hopkins University

A well-studied problem in the streaming model is that of estimating the frequency moments of an underlying vector. Formally, in an *insertion-only stream*, the algorithm is presented with a sequence of integers $\langle a_1, a_2, \ldots, a_m \rangle$ from a universe $[n]$. A *turnstile stream* is defined similarly except integers may also be removed from the stream. The $p$-th frequency moment of the stream is defined to be $F_p = \sum_{i \in [n]} f_i^p$, where $f_i$ is number of times integer $i$ occurs in the stream, i.e., its frequency. The quantity $F_p$ is a basic, yet very important statistic of a dataset (e.g. [AGMS99]). For example, interpreting $0^0$ as 0, $F_0$ is equal to the number of distinct items in the stream. $F_2$ measures the variance and can be used to estimate the size of a self-join in database applications. It also coincides with the (squared) Euclidean norm of a vector and has application in geometric and linear algebraic problems on streams. For other non-integer $p > 0$, $F_p$ can serve as a measure of the entropy or skewness of a dataset, which can be useful for query optimization.

In their seminal paper, Alon, Matias & Szegedy [AMS96] introduced the study of frequency moments in the streaming model. Nearly two decades of research have been devoted to the space and time complexity of this problem. An incomplete list of papers on frequency moments includes [Ind00, CKS03, Woo04, IW05, BO10, KNPW11, And12], and [BKSV14]; please also see the references therein. In the turnstile model, $\Theta(\epsilon^{-2} \log(mn))$ bits of space is necessary and sufficient for a randomized one-pass streaming algorithm to obtain a $(1 \pm \epsilon)$-approximation to $F_p$ for $0 < p \leq 2$ [KNW10]. Here, by $(1 \pm \epsilon)$-approximation, we mean that the algorithm outputs a number $\widetilde{F_p}$ for which $(1 - \epsilon)F_p \leq \widetilde{F_p} \leq (1 + \epsilon)F_p$. For larger values of $p$, i.e., $p > 2$, the memory required becomes polynomial in $n$ rather than logarithmic [CKS03, IW05].

In this paper, we study the frequency moment estimation problem in the *random-order model*, which is a special case of insertion streams in which the elements in the stream occur in a uniformly random order and the algorithm sees the items in one pass over this random order. This model was initially studied by Munro and Paterson [MP80], which was one of the initial papers on data streams in the theory community. Random-order streams occur in many real-world applications and are studied in, e.g., [DLOM02, CJP08, GH09] and the references therein. It has been shown that there is a considerable difference between random order streams and general arbitrary order insertion streams for problems such as quantile estimation [CJP08, GM09]. For other problems, such as frequency moments for $p > 2$, the dependence on $n$ in the space complexity is roughly the same in random and general arbitrary order streams [GH09]. Notice that [CMVW16] studies the frequency moments problem of stochastic streams, in which data points are generated from some distribution. That model is different from ours, but when conditioning on the realizations of the values of each point, the stream is exactly in random order. Therefore, our analysis is applicable to that model as well.

However, there is a gap in our understanding for the important problem of $F_p$-estimation, $0 < p \leq 2$, in the random order model. On the one hand there is an $\Omega(\epsilon^{-2})$ bits of space lower bound [CCM08]. On the other hand, the best upper bound we have is the same as the best upper bound in the turnstile model, namely $O(\epsilon^{-2} \log n)$ bits [AMS96, KNW10, KNPW11]. In practice it would be desirable to obtain $O(\epsilon^{-2} + \log n)$ bits rather than $O(\epsilon^{-2} \log n)$ bits, since if $\epsilon$ is very small this can lead to considerable savings. For example, if $\epsilon^{-2} \approx \log n$, it would represent a roughly quadratic improvement. The goal of this work is to close this gap in our understanding of the memory required for $F_p$-estimation, $0 < p \leq 2$, in the random order model.

## 1.1 Our Contribution

In this paper, we make considerable progress on understanding the space complexity of $F_p$ with $0 < p \leq 2$, in random order streams. Specifically,

- for $F_2$, we show that there exists a simple, and in fact deterministic, one-pass algorithm using space $\mathcal{O}(\epsilon^{-2} + \log n)$ bits to output a $(1 \pm \epsilon)$ approximation, provided $F_2 \geq m \cdot \log n$,

- for $F_p$ with $p \in (0, 2) \backslash \{1\}$, we obtain a one-pass deterministic algorithm to obtain a $(1 \pm \epsilon)$ approximation using $\widetilde{O}(\epsilon^{-2} + \log n)$ bits of space. We also show that this space complexity is optimal for deterministic algorithms in the random order model, up to $\log(1/\epsilon) + \log \log n$ factors. Note that for the case $p = 1$, $F_p$ is the length of the stream, which can be computed in $O(\log n)$ bits of memory.

## 1.2 Our Techniques

For $F_2$, we partition the stream updates into a sequence of small blocks for which each block can be stored using $\mathcal{O}(\epsilon^{-2})$ bits. We then construct an unbiased estimator by counting the number of "pairs" of updates that belong to the same universe item. The counting can be done exactly and with only $\mathcal{O}(\log n)$ bits of space in each small block. We further show that if $F_2 \geq \log n \cdot F_1$, we can obtain the desired concentration by averaging the counts over all blocks. The analysis of the concentration is by constructing a Doob martingale over the pairs and applying Bernstein's inequality.

For $F_p$ with $0 < p < 2$ our algorithm is considerably more involved. We first develop a new reduction from estimating $F_p$ to finding $\ell_p$ heavy hitters (a heavy hitter is an item with frequency comparable to $F_p^{1/p}$). In the rest of this section, we illustrate the high level ideas for obtaining a constant factor approximation to $F_p$. Let $v = (f_1, f_2, \ldots, f_n)$ be the frequency vector of the items. Our reduction is to apply a scaling $X_i$ to each $f_i$, where each $X_i$ is pairwise independently drawn from some distribution. We argue that finding the leading $\ell_p$ heavy hitter of the scaled frequency vector (denoted as $X_{i^*} f_{i^*}$) gives a good estimation to the $F_p$ of the original stream. The distribution of $X_i$ is a so-called $p$-inverse distribution (see Definition 2 for details). This distribution has similar tails as the max-stable distributions used in [And12] or the precision sampling distribution [AKO11]. However, it has different properties better suited to our setting, e.g., we can use pairwise independent variables to do the scaling. In contrast, for max-stable distributions, we have to use fully random hash functions and apply pseudo-random generators, which is problematic in our very low space regime (a related technique called tabulation-based hashing has similar problems). Note that [AKO11] does not require pseudo-random generators, but their precision sampling technique aims to solve a more general family of problems and their distribution has a slightly different form, e.g., the random variable is drawn from a continuous distribution. The $p$-inverse distribution is particularly suited for $p$-norm estimation, and allows for a concise algorithm and improved space complexity in our setting.

Next, we group the coordinates of $v$ into $\Theta(\log n)$ levels by their scalings, i.e., if $2^{-w}F_p < X_i^p \leq 2^{-w+1}F_p$, then $i$ is in level $w$ for some integer $w$. Let $Z_w \subset [n]$ be the set of universe items in level $w$. We observe that if $i^* \in Z_w$ and $X_{i^*}^p f_{i^*}^p \approx F_p$, then

$$f_{i^*}^p = \Omega(2^w).$$

Fortunately, we can also show that in expectation, $i^*$ is an $\ell_p$-heavy hitter of the substream induced by $Z_w$. Our algorithm simply *looks for $i^*$ from $Z_w$ for every $w \in [\log n]$*. One may notice that if we run the search instance in parallel, then there will be a $(\log n)$-factor blowup in space. However, we can show that for random order streams, one can choose a

$$w_0 = \Theta\left(\log \log n\right)$$

such that

1. for all $w > w_0$: the search for $i^*$ can be done in one pass and in sequence for each $w$. This is because $f_{i^*}$ is large (i.e., $\Omega(2^w)$) and a small portion of the random order stream contains sufficient information about $i^*$.

2. for all $w \leq w_0$: with high probability, $|Z_w| = \text{poly} \log n$. We thus do a brute force search for each level $w$ below $w_0$ in parallel. Each search instance uses a small amount of memory because of the small universe size.

The final space overhead becomes a $\text{poly}(w_0)$ factor rather than a $\Theta(\log n)$ factor. This observation is critical to reduce space usage for approximating frequency moments in random order streams and is, to the best of our knowledge, new. For $p \geq 2$, the above claim is no longer true. We leave the exact space complexity for $p \geq 2$ as an open problem.

## 1.3 Roadmap

In Section 2, we introduce some definitions and the $p$-inverse distribution. In Section 3, we present our algorithm for $F_2$. In Section 4, we present our generic framework for approximating $F_p$ as well as our main result. In Section 6, we present the detailed construction of each subroutine used in Section 4, and the details of the main algorithm. In Section 7, we introduce our deterministic algorithm, which uses our randomized algorithm as a subroutine. We also show its optimality in the same section.

# 2 Preliminaries

**Definition 1 (Aggregate Streaming)**
*An* insertion-only stream *(or simply* stream*)* $\mathcal{S} = \langle a_1, a_2, \ldots, a_m \rangle$ *is a sequence of integers, where each $a_i \in [n]$. A* weighted stream $\mathcal{S}' = \langle (a_1, w_1), (a_2, w_2), \ldots, (a_m, w_m) \rangle$ *is a sequence of pairs, where each $a_i \in [n]$ and each $w_m \in \mathbb{R}$. The insertion-only stream $\mathcal{S}$ is a special case of a weighted stream with all the weights being 1. The frequency $f_i(\mathcal{S}')$ of a weighted stream is defined as the sum of all weights of the item $i$. Formally,*

$$f_i(\mathcal{S}') := \sum_{j=1}^{m} \mathbb{I}(a_j = i) w_j.$$

*The frequency vector $V(\mathcal{S}') \in \mathbb{R}^n$ is the vector with the $i$-th coordinate being $f_i(\mathcal{S}')$, for each $i \in [n]$. The $p$-th frequency moment, for $p \geq 0$, is defined as*

$$F_p(\mathcal{S}') := \|V(\mathcal{S}')\|_p^p := \sum_{i=1}^{n} f_i^p(\mathcal{S}').$$

*For time $0 < t_1 \leq t_2 \leq m$, we denote*

$$\mathcal{S}'^{:t_1} := \langle (a_1, w_1), (a_2, w_2), \ldots, (a_{t_1}, w_{t_1}) \rangle \quad \text{and}$$
$$\mathcal{S}'^{t_1:t_2} := \langle (a_{t_1}, w_{t_1}), (a_{t_1+1}, w_{t_1+1}), \ldots, (a_{t_2}, w_{t_2}) \rangle$$

*as sub-streams of $\mathcal{S}'$ from 1 to $t_1$ or from $t_1$ to $t_2$.*

We introduce a discretized version of the $\alpha$-Fréchet distribution: the $\alpha$-Inverse distribution.

**Definition 2** *Fixing an $\alpha > 0$, we say a random variable $X$ on $\mathbb{N}$ is drawn from an $\alpha$-*Inverse *distribution if*
$$\mathbb{P}[X < x] = 1 - \frac{1}{x^\alpha} \quad \text{for} \quad x \in \mathbb{N}_+.$$

**Definition 3 ($\alpha$-Scaling Transformation)**
*Given an $\alpha > 0$, an $\alpha$-scaling transformation (ST) $\mathcal{T}_{k,\alpha} : [n]^m \to ([k] \times [n] \times \mathbb{R})^m$ is a function acting on a stream of length $m$ on universe $[n]$. On input stream $\mathcal{S}$, it outputs a weighted stream $\mathcal{S}'$ of length $km$ on universe $[k] \times [n]$ via the following operation: let $X_{i,j}$ be identical independent (or with limited independence) $\alpha$-Inverse random variables, where $i = 1, 2, \ldots, k$ and $j = 1, 2, \ldots, n$. For each $a \in \mathcal{S}$, the transformation outputs*

$$\mathcal{T}_{k,\alpha}(a) \to ((a,1), X_{1,a}), ((a,2), X_{2,a}), \ldots, ((a,k), X_{k,a}).$$

The next lemma shows that the $\frac{k}{2}$-th largest element of the transformed frequency vector gives a good approximation to the $\alpha$-norm of the vector.

**Lemma 4** *Let $\mathcal{S}$ be a stream of universe $[n]$. Let $\mathcal{T}_{k,\alpha}$ be a pairwise independent $\alpha$-ST with $k \geq \frac{160}{\alpha^2 \epsilon^2}$ being an even integer and $\alpha \geq 0$, where $\epsilon \in (0, \frac{1}{2\alpha})$. Let $\mathcal{S}' = T_{k,\alpha}(\mathcal{S})$. Define the two sets,*

$$U_+ := \left\{ (a,r) \in [n] \times [k] : f_{(a,r)}(\mathcal{S}') \geq 2^{1/\alpha}(1-\epsilon) \|V(\mathcal{S})\|_\alpha \right\}$$

and

$$U_- := \left\{(a,r) \in [n] \times [k] : f_{(a,r)}(\mathcal{S}') < 2^{1/\alpha}(1+\epsilon) \left\| V(\mathcal{S}) \right\|_\alpha \right\}.$$

**Then with probability at least** $0.9$,

$$|U_+| \geq \tfrac{k}{2}, \quad |\overline{U_-}| < \tfrac{k}{2} \quad \text{and} \quad |U_+ \cap U_-| > 0,$$

where $\overline{U}$ is the complement of the set $U$.

The proof of this lemma is provided in Section A.

## 3 A Simple $F_2$ Algorithm For Long Streams

We start with a very simple algorithm for approximating $F_2$ in a random order stream. We denote the algorithm as `RANDF2`. The algorithm has a parameter $b > 0$. It treats the stream updates as a series of length $b$ blocks, i.e.,

$$\mathcal{S} = (B_1, B_2, \ldots, B_{m/b}).$$

Initialize a register $K = 0$. At any time, suppose the current block is $B_i$. The algorithm simply stores everything it sees until the end of $B_i$. After storing $B_i$ entirely, the algorithm computes

$$K_i = \sum_{j=1}^{} \binom{f_j(B_i)}{2}.$$

This can be done using space $b \log n$ bits. Then, the counter $K$ is updated as

$$K \leftarrow K + K_i.$$

At the end of the stream, the algorithm computes

$$Y = \frac{2K(m^2 - m)}{(b^2 - b)T} + m,$$

where $T$ is the ID of the last complete block, and $m$ is the length of the stream. The algorithm uses $O(b \log n)$ bits. By setting

$$b = \Theta\left[\max\left(\tfrac{1}{\epsilon^2 \log n}, 2\right) \cdot \log \tfrac{1}{\delta}\right].$$

we obtain the following theorem.

**Theorem 5** *Let $\mathcal{S}$ be a random order stream satisfying $F_2(\mathcal{S}) \geq m \cdot \log n$. After one pass over the stream $\mathcal{S}$, $Y$ is a $(1 \pm \epsilon)$ approximation to $F_2(\mathcal{S})$ with probability at least $1 - \delta$. Moreover, to compute $Y$, `RANDF2` uses $\mathcal{O}\left(\epsilon^{-2} \log \delta^{-1} + \log n\right)$ bits of memory.*

**Proof** For each $j \in [n]$, we distinguish its stream updates as $u_1^{(j)}, u_2^{(j)}, \ldots, u_{f_j(\mathcal{S})}^{(j)}$. Then $K_i$ is the number of pairs of the form $(u_{\ell_1}^{(j)}, u_{\ell_2}^{(j)})$ appearing in $B_i$. We denote $F_2 = F_2(\mathcal{S})$ and $F_1 = F_1(\mathcal{S})$ for simplicity. Thus,

$$\forall i \in [T] : \mathbb{E}(K_i) = \frac{\sum_{j=1}^n \binom{f_j(\mathcal{S})}{2}}{\binom{F_1}{2}} \binom{b}{2} = \frac{F_2 - m}{m^2 - m} \frac{(b^2 - b)}{2}.$$

Thus, by linearity of expectation,
$$\mathbb{E}[Y] = F_2.$$

We now prove that $Y$ is concentrated around its mean. Notice that the algorithm can be viewed as sampling a number of "pairs". A pair is formed by two updates to the same universe element. There are $q = (F_2 - F_1)/2$ many pairs. Let $P = [q]$ denote the set of pairs. For each pair $z \in P$, we let $X_z$ denote the indicator that $X_z$ is sampled by some bucket. Let $K = \sum_{z \in P} X_z$. Note that this $K$ is the same as the one denoted in the algorithm. Let $Q_z = \mathbb{E}[K | X_1, X_2, \ldots, X_z]$. Then $Q_z$ for $z = 1, 2, \ldots$ form a Doob martingale. Also notice that $|Q_z - Q_{z-1}| \leq 1$. Next, we proceed to bound the variance of

$$Q_z - Q_{z-1} = X_z | X_1, X_2, \ldots X_{z-1}.$$

For a pair $z \in P$, let $a, b$ be the two nodes. Consider a fixed assignment of $X_1, X_2, \ldots X_{z-1}$. Also note that, knowing $X_i$, the two nodes of the $i$-th pair are assigned to some block.

Now, if from the information of $X_1, X_2, \ldots, X_{z-1}$, $a, b$ are both assigned and to the same block, then $X_z = 1$ and otherwise $X_z = 0$. For both cases $\mathbb{V}(Q_z - Q_{z-1}) = 0$. If $a, b$ are assigned, but it cannot be determined if they are in the same block, then $\mathbb{P}[X_z = 1] \leq b/m$ and thus $\mathbb{V}(X_z) \leq b/m$. If only one of $a, b$ is assigned, then $\mathbb{P}[X_z = 1] \leq b/m$, and thus $\mathbb{V}(X_z) \leq b/m$. Lastly, if both $a, b$ are not assigned, then $\mathbb{P}[X_z = 1] \leq b/m$. Thus $\mathbb{V}(X_z) \leq b/m$. Overall, we have that $v_z^2 := \mathbb{V}(Q_z - Q_{z-1}) \leq b/m$ for all possible $X_1, X_2, \ldots, X_{z-1}$. Let

$$V = \sum_z v_z^2 \leq \frac{b(F_2 - m)}{2m}.$$

Next by Bernstein's inequality [DP09], we have that,

$$\mathbb{P}[|K - \mathbb{E}(K)| \geq t] \leq 2 \exp\left(-\frac{t^2}{2V(1 + t/3V)}\right).$$

Since we need to have a $(1 \pm \epsilon)$ approximation to $F_2$, we can set

$$\frac{2t(m^2 - m)b}{(b^2 - b)m} \leq \epsilon F_2 \quad \text{and} \quad t = \epsilon \frac{F_2 b}{2m}.$$

Since $F_2 \geq m \cdot \log n \cdot \log \frac{1}{\delta}$, and $\epsilon$ is sufficiently small, we can bound the error probability by:

$$\mathbb{P}[|K - \mathbb{E}(K)| \geq t] \leq 2 \exp\left(-\frac{t^2}{2V(1 + t/3V)}\right)$$
$$\leq 2 \exp\left(-\frac{\epsilon^2 F_2^2 b}{8m(F_2 - m)}\right) \leq \delta, \quad (1)$$

for $b = \Omega\left(\frac{\log \frac{1}{\delta}}{\epsilon^2 \cdot \log n}\right)$. Finally, since $K \in \mathbb{E}(K) \pm \epsilon F_2 b/(2m)$, we have

$$Y \in F_2 \pm \epsilon \frac{F_2 b}{2m} \cdot \frac{2(m^2 - m)b}{(b^2 - b)m} \subset (1 \pm \epsilon) F_2$$

as desired. ∎

## 4 A Generic Framework for $F_p$ Estimation

In this section, we first construct a generic framework for $F_p$ estimation, and then we construct all the components in subsequent sections. For a random order stream $\mathcal{S}$, we will need the following three components to construct an $F_p$ estimation algorithm.

- A counter that stores the time for the current update;
- An algorithm that gives a constant approximation to the current $F_p(\mathcal{S}^{:t})$;
- An algorithm that computes an accurate $(1 \pm \epsilon)$ approximation to $F_p(\mathcal{S})$ given $\text{poly}(\log n, \epsilon^{-1})$ approximations to $m$ and $F_p(\mathcal{S})$.

To begin, we denote by C2Fp a data structure that, once initialized with a $\text{poly}(\log n, \epsilon^{-1})$-approximation of both the length and $p$-th frequency moments, $F_p(\mathcal{S})$, supports two operations: update and query. At the end of the stream, C2Fp.query() returns either Fail or a $(1 \pm \epsilon)$-approximation to $F_p$ of the input stream $\mathcal{S}$. Component (1) will be used to guess the length of the stream, and component (2) will be used to guess an approximation to $F_p(\mathcal{S})$. We denote component (2) by ConstFp, which is a data structure that supports both update and query operations. ConstFp.query() returns a 2-approximation to $F_p(\mathcal{S}^{:t})$ at some fixed $t$. The full framework is described in Algorithm 1.

Our full algorithm is denoted by RndF$_p$, which uses C2Fp as a subroutine. From a high level, the algorithm constantly guesses the length of the stream. If at some point in time the algorithm finds that the current guess of the length is at least a factor $C = \text{poly}(\epsilon^{-1}, \log n)$ smaller than the true value of the stream, then the algorithm initializes a new instance of C2Fp to estimate the $F_p$ value of the stream. At the end, it is guaranteed that a stored instance of C2Fp uses at least a $(1 - \text{poly}(\epsilon, 1/\log n))$ portion of the stream to approximate the frequency moments. It can be verified that an accurate estimation of $F_p$ of this portion of the stream will serve as a good estimator for the overall stream. Therefore, if C2Fp is able to output the correct answer with high probability, then the algorithm RndF is guaranteed to be correct with high probability.

**Theorem 6 (Main Theorem)** *For fixed $p \in [0, 2)$, $\epsilon \in (0, 1)$, $\delta \in (\frac{1}{\text{poly}(n)}, \frac{1}{2})$, and $n \in \mathbb{N}_+$, algorithm RndF$_p$, making a single pass over a random order stream $\mathcal{S}$ on universe $[n]$, outputs a number $\widehat{F}$ such that, with probability at least $1 - \delta$,*

$$(1 - \epsilon) F_p(\mathcal{S}) \leq \widehat{F} \leq (1 + \epsilon) F_p(\mathcal{S}),$$

---
**Algorithm 1:** Full algorithm for $F_p$ in random order: $\text{RndF}_p$
---
**Data:**
  $\mathcal{S} = \langle a_1, a_2, \ldots, a_m \rangle$ is random order stream of length $m$ from universe $[n]$ (known in advance);
  $p \in [1, 2]$, a real constant;

1 **Initialize** $(p, n, \epsilon, \delta)$:
2    $m_0 \leftarrow 1$, $m_1 \leftarrow 1$ $G_0 \leftarrow 1$. Here $m_0$ is the approximated length, $G_0$ is a guess of $F_p$, $\epsilon$ is the target precision, and $\delta$ is the failure probability;
3    $A_1 \leftarrow$ new $\text{C2Fp}(p, \epsilon/3, n, m_0, G_0, \delta/3)$, $A_2 \leftarrow$ new $\text{C2Fp}(p, \epsilon/3, n, m_0, G_0, \delta/3)$;
4    $A_3 \leftarrow$ new $\text{ConstFp}(p, n, \delta/3)$;
5    $C \leftarrow \text{poly}(\frac{1}{\epsilon}, \log \frac{n}{\delta})$;
6 **Update** $a$:
7    $A_1.\text{update}(a)$, $A_2.\text{update}(a)$, $A_3.\text{update}(a)$;
8    $m_1 \leftarrow m_1 + 1$;
9    **if** $m_1 \geq Cm_0$ **then**
10      $A_1 \leftarrow A_2$;
11      $G_0 \leftarrow A_3.\text{query}()$;
12      $m_0 \leftarrow m_1$;
13      $A_2 \leftarrow$ new $\text{C2Fp}(p, \epsilon/3, n, m_0, G_0, \delta/3)$;
14 **Query()**:
15    return $A_1.\text{query}()$;

---

*where the probability is over both the randomness of the algorithm and the randomness of the data stream. The algorithm uses*

$$\mathcal{O}\left[\left(\frac{1}{\epsilon^2}\left(\log\log n + \log\frac{1}{\epsilon}\right)^4 + \log n\right)\log\frac{1}{\delta}\right]$$

*bits of memory in the worst case.*

**Proof** Without loss of generality, we assume $F_p(\mathcal{S}) = \Omega\left(\text{poly}\frac{1}{\epsilon}\text{poly}\log\frac{n}{\delta}\right)$, since otherwise we can use a turnstile $F_p$ algorithm with memory $\mathcal{O}\left(\frac{1}{\epsilon^2}\log\log\frac{n}{\delta} + \frac{1}{\epsilon^2}\log\frac{1}{\epsilon}\right)$ bits to solve the $F_p$ estimation problem. Initialize $m_0 = 1$. Let $\mathcal{S}' = \mathcal{S}^{m_0:m}$. By definition of the algorithm, $A_1$ is an instance of $\text{C2Fp}$ that runs on $\mathcal{S}'$. Let $\mathcal{S}'' = \mathcal{S}^{0:m_0}$ and $C = \text{poly}\,\epsilon^{-1}\,\text{polylog}\,\frac{n}{\delta}$. By definition of the algorithm, we always update $m_0$ such that

$$\frac{m}{C^2} \leq m_0 \leq \frac{m}{C},$$

at the end of the stream. By Lemma 25, and Lemma 26 with probability at least $1 - \delta/3$,

$$\frac{F_p(\mathcal{S})}{5^p C^{2p}} \leq F_p(\mathcal{S}'')$$
$$\leq \left(17 \log \frac{20n}{\delta}\right)^p \left(C^{-1} F_p(\mathcal{S}) + 4 \log \frac{20n}{\delta}\right)$$
$$\leq \frac{\left(18 \log \frac{20n}{\delta}\right)^p}{C} F_p(\mathcal{S}),$$

where the last inequality holds for sufficiently large $F_p(\mathcal{S})$. Conditioned on this event, we obtain that

$$\|V(\mathcal{S}'')\|_p \leq \frac{\left(18 \log \frac{20n}{\delta}\right)}{C^{1/p}} \|V(\mathcal{S}))\|_p \leq \frac{\epsilon}{3p} \|V(\mathcal{S}))\|_p,$$

for sufficiently large $C$. By the triangle inequality, we obtain

$$(1 - \epsilon/(3p)) \|V(\mathcal{S})\|_p \leq \|V(\mathcal{S})\|_p - \|V(\mathcal{S}'')\|_p$$
$$\leq \|V(\mathcal{S}')\|_p \leq \|V(\mathcal{S})\|_p.$$

Thus $F_p(\mathcal{S}')$ is a $(1 \pm \epsilon/3)$ approximation to $F_p(\mathcal{S})$.

In the algorithm, $A_3$ is an instance of `ConstFp`, i.e., by Theorem 7. Let $G_0$ be the output of $A_3$.query(). Since with probability at least $1 - \delta/3$, $A_3$ outputs a $c_0$ approximation to $F_p(\mathcal{S}'')$ for some constant $c_0$, we obtain that $G_0$ is a

$$5^p c_0 C^{2p} \left(\log \frac{20n}{\delta}\right)^{p-1}$$

approximation to $F_p(\mathcal{S}')$.

In the algorithm, $A_1$ is an instance of `C2Fp`, which runs on the stream $\mathcal{S}'$ at any time $t$ with the required parameters, i.e., $G_0$. By Theorem 8, with probability at least $1 - \delta/3$, $A_1$.query() outputs a $(1 \pm \epsilon/3)$ approximation to $F_p(\mathcal{S}')$, and thus a $(1 \pm \epsilon)$ approximation to $F_p(\mathcal{S})$. By a union bound, the overall algorithm is correct with probability at least $1 - \delta$.

By Theorem 16 and Theorem 17, the space needed for $A_1$ and $A_2$ is

$$\mathcal{O}\left[\left(\frac{1}{\epsilon^2}\left(\log\log n + \log \frac{1}{\epsilon}\right)^4 + \log n\right) \log \frac{1}{\delta}\right].$$

The space needed for $A_3$ is $\mathcal{O}(\log n \log \frac{1}{\delta})$ (Theorem 7). Thus the total space is dominated by the total space used by $A_1$ and $A_2$ as desired. ∎

The following is a theorem required in the above proof.

**Theorem 7 (Const. $F_p$ approx., [KNW10])**
*For a fixed $n$, there exists a turnstile streaming algorithm, which on input a stream $\mathcal{S}$ of length $m$, outputs a number $F \in (1 \pm \epsilon) F_p(\mathcal{S})$ with probability at least $1 - \delta$. The algorithm uses $\mathcal{O}(\epsilon^{-2} \log m + \log \log(n)) \log \delta^{-1})$ bits of space in the worst case.*

In subsequent sections, we will construct the `C2Fp` Algorithm.

# 5  A $(1 \pm \epsilon)$ Approximation to $F_p$ With a Prior

In this section, we construct the algorithm C2Fp. We assume that the input is a random order stream and that the algorithm is given two parameters, $\widehat{m}$ and $G$, which are $\text{poly}(\epsilon^{-1}, \log n)$ approximations to the length and the $p$-th frequency moments of the stream $\mathcal{S}$, respectively.

## 5.1  High Level Idea

Although the high level idea is introduced in the introduction, we repeat it here with more details for better understanding of the algorithm. To illustrate the intuition of the algorithm, we first consider a constant factor approximation algorithm. Estimating the frequency moments can be reduced to finding the *heavy hitters* of a scaled vector, as shown in Lemma 4. Suppose the frequency vector in a stream is $v = (f_1(\mathcal{S}), f_2(\mathcal{S}), \ldots, f_n(\mathcal{S}))$, and the scaling applied to it is $X = (X_1, X_2, \ldots, X_n)$, where the $X_i$ are pairwise independent $p$-Inverse (see Definition 2) random variables. Let $i^*$ be the maximum of the scaled vector. By Lemma 4, we expect
$$X_{i^*}^p f_{i^*}^p(\mathcal{S}) \approx F_p.$$
We group the coordinates of $v$ into $\Theta(\log n)$ levels by their scalings, i.e., if $2^{-w} F_p < X_i^p \leq 2^{-w+1} F_p$, then $i$ is in level $w$. Let $Z_w \subset [n]$ be the universe items of level $w$. We observe that if $i^* \in Z_w$, then
$$f_{i^*}^p(\mathcal{S}) = \Omega(2^w).$$
Luckily, we can also show that, in expectation, $i^*$ is an $F_p$-heavy hitter of the substream induced by $Z_w$. Our algorithm is simply *looking for $i^*$ from $Z_w$ for every $w \in [\log n]$*. One may notice that if we run the search instance in parallel, then there will be a $\log n$ factor blowup in the space. However, we can show that in a random order stream, one can choose an $w_0 = \Theta(\log \log n)$ such that

1. for all $w > w_0$: the search for $i^*$ can be done in one pass and in sequence for each $w$.

2. for all $w \leq w_0$: with high probability, $|Z_w| = \text{poly} \log n$. We thus do a brute force search for each level $w$ below $w_0$ in parallel.

The final space overhead is a $\text{poly}(w_0)$ factor rather than $\Theta(\log n)$.

To reduce the error from constant to $(1 \pm \epsilon)$, we repeat the above process $\Theta\left(\frac{1}{\epsilon^2}\right)$ times conceptually. Namely, we apply a $p$-ST $\mathcal{T}_{k,p}$ transformation to the stream, where $k = \Theta\left(\frac{1}{\epsilon^2}\right)$. For $r = 1, 2, \ldots, [k]$, $i = 1, 2, \ldots, n$, we denote the scaling $p$-Inverse random variable as $X_i^{(r)}$. We wish to find the heavy hitter for each $r$ using the same procedure described above. By Lemma 4, the $k/2$-th largest of all the outputs serves as a good approximation to $F_p(\mathcal{S})$.

## 5.2  The Algorithm

The algorithm needs three components, SmallApprox, SmallCont and LargeCont. All these algorithms support "update" and "query" operations. SmallApprox returns fail if $F_p(\mathcal{S})$ is much larger than $\operatorname{poly}(\epsilon^{-1}, \log n)$, otherwise returns an approximation to $F_p(\mathcal{S})$. `SmallApprox` is a turnstile streaming $F_p$ algorithm [KNW10] but with restricted memory. Once the memory exceeds the memory quota, the algorithm simply returns `Fail`. `SmallCont` estimates the contribution from the small-valued frequencies and `LargeCont` estimates the contribution from the large-valued frequencies. The correctness of these algorithms is presented in Theorem 17, and 16. The full algorithm is presented in Algorithm 2. The following theorem guarantees its correctness.

**Theorem 8** *Fix $p \in [0, 2]$, $\epsilon \in (0, 1/2)$ and $\delta = \Omega(1/\operatorname{poly}(n))$. Let $\mathcal{S}$ be a random order stream on universe $[n]$ and with length $m$. Given that $C_0^{-1} F_p(\mathcal{S}) \leq G_0 \leq F_p(\mathcal{S})$ and $C_0^{-1} m \leq m_0 \leq m$ for some $C_0 = \operatorname{poly}(\epsilon^{-1}, \log n)$, there exists an algorithm $\mathcal{A}$, which makes a single pass over $\mathcal{S}$ and outputs a number $F$ such that*

$$F \in (1 \pm \epsilon) F_p(\mathcal{S})$$

*with probability at least $1 - \delta$, where the probability is over both the randomness of the algorithm and of the order of the stream. The algorithm uses*

$$\mathcal{O}\left[\left(\tfrac{1}{\epsilon^2}\left(\log\log n + \log\tfrac{1}{\epsilon}\right)^4 + \log n\right)\log\tfrac{1}{\delta}\right]$$

*bits of memory in the worst case.*

We postpone the full proof and detailed algorithmic constructions to the appendix.

## 6 Continue on $(1 \pm \epsilon)$ Approximation to $F_p$ With a Prior

**Proof** [Proof of Theorem 8] The proof relies on the correctness of `SmallApprox`, `SmallCont` and `LargeCont`, which is guaranteed by Theorem 16 and Theorem 17 and Lemma 7. We prove the theorem by setting $\delta = 0.9$. For general $\delta = \Omega(1/\operatorname{poly}(n))$, one only needs to repeat the Algorithm 2 in parallel $\mathcal{O}(\log \tfrac{1}{\delta})$ times and output the median of theses instances. Notice that the parallel repetitions rely on the same stream, which makes the repetitions not independent. We handle this issue at the end of the proof.

By the correctness of `SmallApprox`, if it does not return `Fail`, then it returns a number $F \in (1 \pm \epsilon) F_p(\mathcal{S})$ with probability at least $0.99$. `SmallApprox` uses space

$$\mathcal{O}\left[\frac{1}{\epsilon^2}\left(\log\log n + \log\frac{1}{\epsilon}\right)\log\frac{1}{\delta}\right].$$

If $m \leq \operatorname{poly}(\epsilon^{-1}, \log n)$, then `SmallApprox` never returns `Fail`.

We now suppose $m \geq \operatorname{poly}(\epsilon^{-1}, \log n)$. By Theorem 17, $B_3$, the instance of `SmallCont`, with probability at least $0.97$, returns a sequence of numbers which are approximations to the top $k$ $p$-Inversely scaled frequencies of $\mathcal{S}$ for large enough $X_i^{(r)}$s. Similarly for $B_2$, except this is for the small $X_i^{(r)}$s. Let $L = B_1.\operatorname{query}() \cup B_2.\operatorname{query}()$ be the sequence returned. By Lemma 4,

**Algorithm 2:** $F_p$-Algorithm with Approximation: $\texttt{C2Fp}(p, \epsilon, n, m_0, G_0)$

**Data:**
    $p \in [0, 2]$, a real number;
    $L \in \left[\frac{F_p(\mathcal{S})}{C_0}, F_p(\mathcal{S})\right]$ for some $C_0 = \Theta\left[\text{poly}(\epsilon^{-1}, \log n)\right]$;
    $m_0 \in \left[\frac{m}{C_0}, m\right]$;
    $\mathcal{S} = \langle a_1, a_2, \ldots, a_m \rangle$ is random order stream of length $m$;
**Result:** $F \in (1 \pm \epsilon) F_p(\mathcal{S})$;

1 **Initialize**$(p, n, \epsilon, \delta)$:
2     $k \leftarrow \Theta\left(\frac{1}{\epsilon^2}\right)$;
3     $X_i^{(r)} \sim p\text{-Inverse distribution for } i \in [n] \text{ and } r \in [k]$, pairwise independent;
4     $w_0 \leftarrow d_0 \left(\log \log n + \log \frac{1}{\epsilon}\right)$ for some large constant $d_0$;
5     Denote $K \in \mathbb{R}^{n \times k}$, whose coordinate $K_{i,r} = X_i^{(r)} v_i$ (only for notational purposes);
6     $B_1 \leftarrow$ new $\texttt{SmallApprox}(p, n, \epsilon)$;
7     $B_2 \leftarrow$ new $\texttt{SmallCont}(p, n, k, \epsilon, w_0, L, \{X_i^r\})$;
8     $B_3 \leftarrow$ new $\texttt{LargeCont}(p, n, k, \epsilon, w_0, L, \{X_i^r\})$;
9 **Update**$(a)$:
10     $B_1.\text{update}(a)$; $B_2.\text{update}(a)$; $B_3.\text{update}(a)$;
11 **Query:**
12     if $B_1.\text{query}() \neq \texttt{Fail}$ then
13         return $B_1.\text{query}()$
14     else if $B_2.\text{query}() = \texttt{Fail}$ or $B_3.\text{query}() = \texttt{Fail}$ then
15         return Fail;
16     else
17         $R \leftarrow$ the $(k/2)$-th largest element of $B_2.\text{query}() \circ B_3.\text{query}()$;
18         return $(R)^p / 2$

with probability at least 0.99, the $(k/2)$-th largest element of $L$ is in $(1 \pm \epsilon) 2^{1/p} F_p^{1/p}(\mathcal{S})$. After a proper scaling, we obtain the desired approximation.

The space requirement follows from Theorem 17 and Theorem 16.

We now show why the median of repetitions boosts the probability from 0.9 to an arbitrary $1 - \delta$. Let $\Omega_1$ be the space of all possible random bits of the pairwise-independent $p$-ST. Since it is pairwise independent, we have that $|\Omega_1| = \text{poly}(n)$. Additionally, let $\Omega_2$ be the set of all possible orders of the stream and $\Omega_3$ be the set of all other random bits used in the algorithm. Consider the events $\mathcal{E}_1$ and $\mathcal{D}_1$ for $\texttt{LargeCont}$ as defined in Theorem 16. We have that $\mathcal{E}_1 = \Omega_1' \times \Omega_2 \times \Omega_3$, where $\Omega_1' \subset \Omega_1$ and $|\Omega_1'|/|\Omega_1| \geq 0.98$. Since $\mathbb{P}[\mathcal{D}_1 | \mathcal{E}_1] \geq 1 - 1/\text{poly}(n)$, we argue that there exists a product space $\mathcal{D}_1' = \Omega_1' \times \Omega_u \subset \mathcal{D}_1$ with

$$|\Omega_u| \geq \left(1 - \frac{1}{\text{poly}(n)}\right) |\Omega_2 \times \Omega_3|.$$

Indeed, for each $\omega \in \Omega_1'$, let $\mathcal{D}_\omega \subset \Omega_2 \times \Omega_3$ be the event $(\mathcal{D}|\omega)$. By the proof of Lemma 16, we have $\frac{|\mathcal{D}_\omega|}{|\Omega_2 \times \Omega_3|} \geq 1 - 1/\operatorname{poly}(n)$. We can choose our parameters such that

$$\forall \omega \in \Omega_1' : |\Omega_1| \left(1 - \mathbb{P}\left[\mathcal{D}_\omega\right]\right) = \frac{1}{\operatorname{poly}(n)}.$$

Therefore, letting $\Omega_u = \cap_{\omega \in \Omega_1'} \mathcal{D}_\omega$, we obtain the desired $\Omega_u$. Hence, conditioned on $\mathcal{D}_1'$, we have $\mathbb{P}[\mathcal{E}_1|\mathcal{D}_1'] \geq 0.98$ and $\mathbb{P}[\mathcal{D}_1'] \geq 1 - \frac{1}{\operatorname{poly}(n)}$.

Similarly, for SmallCont, we can find an event $\mathcal{D}_2'$ such that $\mathbb{P}[\mathcal{D}_2'] \geq 1 - \frac{1}{\operatorname{poly}(n)}$ and $\mathbb{P}[\mathcal{E}_2 \cap \mathcal{F}_2|\mathcal{D}_2'] \geq 0.98$. Now conditioning on event $\mathcal{D}_1' \cap \mathcal{D}_2'$, the algorithm outputs the desired answer with probability at least 0.94. By repeating $\mathcal{O}(\log \delta^{-1})$ times and take the median of all outputs, by a standard argument we boost the probability to $1 - \delta$. ∎

In the following sections we will construct SmallCont in Section 6.2 and LargeCont Section 6.3. Before that, we first introduce our heavy hitter algorithm in random order streams in the next section.

## 6.1 An $F_p$ Heavy Hitters Algorithm in Random Order Streams

In this section, we describe an $F_p$ heavy hitters algorithm for $0 \leq p < 2$ in the random order model. This algorithm serves as a basic routine for approximating the frequency moments. The algorithm outputs the identity of a constant $F_p$-heavy hitter using only a small portion of the stream. The algorithm requires an approximation to $F_p$ and to $F_1$. Formally, the algorithm is presented in Algorithm 8 of Section C. At a high level, if there exists a heavy hitter in the stream, the updates of the heavy hitter are evenly distributed among the stream updates. This property of the random order model enables the algorithm to uses a small trunk of updates to learn the index of the heavy hitter. For the small trunk of stream updates, we use BPTree[BCI+16] to learn the identity of the heavy hitter. The formal guarantee of the BPTree algorithm is as follows.

**Theorem 9 ([BCI+16])** *Given an insertion-only stream $\mathcal{S}$ over a universe $[n]$ with length $m$, after one pass over the stream, algorithm BPTree($\epsilon, \delta$), for some constants $\epsilon, \delta \in (0, 1)$, outputs a set $M \subset [n] \times \mathbb{R}$, such that with probability at least $1 - \delta$, for all $i \in [n]$, if*

$$f_i^2(\mathcal{S}) \geq \epsilon^2 F_2(\mathcal{S})$$

*then $(i, \widehat{f_i}) \in M$ with that $\widehat{f_i} \in (1 \pm \epsilon)f_i(\mathcal{S})$. BPTree uses $\mathcal{O}\left(\frac{1}{\epsilon^2} \log \frac{1}{\delta \epsilon} (\log n + \log m)\right)$ bits of space in the worst case.*

To boost the probability of detecting the heavy hitter, we use $\Theta(\log n)$ sequential trunks of the stream to detect the heavy hitter. We argue that if $i^*$ is a heavy hitter, then with high probability $i^*$ is detected in a constant fraction of these trunks. We use the Misra-Gries [MG82] algorithm, MG, to store heavy hitter indices. MG is a deterministic algorithm. The formal guarantee of this algorithm is,

**Theorem 10 ([MG82])** *Given a stream $\mathcal{S}$ over a universe $[n]$ with length $m$, after one pass over the stream, MG($c$), for some constant $0 < c \leq 1$, outputs a set $M \subset [n]$, such that if*

$$f_i(\mathcal{S}) \geq cm$$

*then $i \in M$. MG uses $\mathcal{O}(\log n)$ bits of space in the worst case.*

To show the guarantees of the algorithm, we first show the following lemma.

**Lemma 11** *Let $\mathcal{S} = \langle a_1, a_2, \ldots, a_m \rangle$ be a random order stream with universe $[n]$. Let $\mathcal{S}' = \langle a_{i_1}, a_{i_2}, \ldots, a_{i_k} \rangle$ be a substream of a fixed sequence of distinct integers $\{i_1, i_2, \ldots, i_k\} \subset [m]$. Then*

$$\mathbb{E}\left[F_2\left(\mathcal{S}'\right)\right] \leq k + \frac{k^2}{m^2} \cdot F_2(\mathcal{S}).$$

**Proof** We put a distinct label on each aggregate of each item. Let $K_{ij}$ be the indicator of the $j$-th update of an item $i$ being included in $\mathcal{S}'$. Let

$$X_i = \left(\sum_{w=1}^{f_i(\mathcal{S})} K_{il}\right)^2.$$

We have

$$X_i = \sum_{w=1}^{f_i(\mathcal{S})} K_{il} + 2 \sum_{w_1 < l_2} K_{il_1} K_{il_2}$$

and

$$\mathbb{E}\left(K_{il_1} K_{il_2}\right) = \frac{k(k-1)}{m(m-1)} \leq \frac{k^2}{m^2}.$$

Additionally with $\mathbb{E}\left(K_{i,l}\right) = \frac{k}{m}$, we obtain the desired statement. ∎

**Lemma 12** *Let $\mathcal{S} = \langle a_1, a_2, \ldots, a_m \rangle$ be a random order stream with universe $[n]$. Let $S_1, S_2, \ldots, S_t$ be distinct substreams of $\mathcal{S}$ (no overlaps), each of length $k$. Suppose $2tk \leq m$ and $t = \Theta(\log n)$. Then, with probability at least $1 - \frac{1}{\text{poly}(n)}$, $\exists I \subset [t]$ s.t. $|I| \geq 0.99t$, and for each $i \in I$,*

$$F_2(S_i) \leq c_4 \cdot \left(k + \frac{4k^2}{m^2} F_2(\mathcal{S})\right)$$

*for some positive absolute constant $c_4$.*

**Proof** For each $i \in [t]$, let $K_i$ be the indicator that

$$F_2\left(S_i\right) \leq c_4 \cdot \left(k + \frac{4k^2}{m^2} \cdot F_2(\mathcal{S})\right).$$

Consider the Doob martingale formed by $\mathbb{E}(\sum K_j | K_1, K_2, \ldots, K_i)_{i=1}^t$. Let

$$W_i = \mathbb{E}\left[\sum K_j | K_1, K_2, \ldots, K_i\right]$$
$$- \mathbb{E}\left[\sum K_j | K_1, K_2, \ldots, K_{i-1}\right].$$

We observe that,

$$\mathbb{E}\left[F_2(\mathcal{S}_i) | K_1, K_2, \ldots, K_{i-1}\right] \le k + \frac{k^2}{(m-tk)^2} F_2(\mathcal{S}')$$
$$\le k + \frac{4k^2}{m^2} F_2(\mathcal{S}).$$

where $\mathcal{S}'$ is the stream after removing the initial $i-1$ trunks. By Markov's inequality, we have

$$\mathbb{P}\left(F_2(S_i) > c_4 \cdot \left(k + \frac{4k^2}{m^2} \cdot F_2(\mathcal{S})\right)\right) \le 0.001,$$

for some constant $c_4$. Hence $\mathbb{E}(\sum K_j) \ge 0.999t$. By the Azuma-Hoeffding inequality,

$$\mathbb{P}\left(\left|\sum K_j - \mathbb{E}\left(\sum K_j\right)\right| \le 0.001t\right) \le e^{-\Omega(t)} \le \frac{1}{\text{poly}(n)},$$

as desired. ∎

**Theorem 13** *Let $\mathcal{S} = \langle a_1, a_2, \ldots, a_m \rangle$ be a random order stream. Suppose*

$$F_p(\mathcal{S}) = \Omega\left[\text{poly}\left(\tfrac{\log n}{\epsilon}\right)\right] m \lesssim \widehat{m} \le \text{poly}\left(\tfrac{\log n}{\epsilon}\right) m.$$

*Then, conditioned on an event $\mathcal{E}$ with probability at least $1 - \frac{1}{\text{poly}(n)}$, algorithm* `HHR` *returns a set $H \subset [n]$ such that if $f_i^p \ge c_0 F_p(\mathcal{S})$ then $i \in H$ and $|H| = \Theta(1)$. The algorithm halts with at most*

$$\mathcal{O}\left[\frac{\text{poly}(\epsilon^{-1}, \log n)\, \widehat{m}^2}{F_p^{2/p}(\mathcal{S})}\right]$$

*stream updates.*

**Proof** Throughout the proof, we use $\text{poly}(n)$ to denote some large enough polynomial in $n$. Since $F_p(\mathcal{S}) = \Omega\left[\text{poly}\left(\tfrac{\log n}{\epsilon}\right)\right]$, we have that the number of stream updates consumed by the algorithm is

$$\log n \cdot \frac{\widehat{m}^2}{F_p^{2/p}} \le \left(\log n \cdot \frac{\text{poly}(\epsilon^{-1} \log n)\widehat{m}}{F_p^{2/p}}\right) m \ll m,$$

for sufficiently large $F_p$. Therefore, the algorithm will not output `Fail`. Let $\mathcal{A}$ be the event that $\exists I \subset [t]$ with $|I| \ge 0.99t$ and for each $i \in I$,

$$F_2(\mathcal{S}_i) \le c_5\left(m_1 + \frac{m_1^2}{m^2} F_2(\mathcal{S})\right)$$

for some constant $c_5$. By Lemma 12,
$$\mathbb{P}[\mathcal{A}] \geq 1 - \frac{1}{\text{poly}(n)},$$
for some large polynomial in $n$. Let $H^* \subset [n]$ be the set of $(c_0, F_p)$-heavy hitters in $\mathcal{S}$, i.e., $i \in H^*$ if and only if $f_i^p(\mathcal{S}) \geq c_0 F_p(\mathcal{S})$. It is clear that $|H^*| = \Theta(1)$. For each $j \in H^*$, let $\mathcal{B}_j^i$ be the event that
$$f_j(\mathcal{S}_i) \geq c_6 \cdot \frac{m_1}{m} \cdot f_j(\mathcal{S}) \geq \frac{c_0^{1/p} m_1}{m} \cdot F_p^{1/p}(\mathcal{S}).$$

Moreover, we have
$$\frac{m_1}{m} \cdot F_p^{1/p}(\mathcal{S}) \geq \frac{m_1}{m} \sqrt{F_2(\mathcal{S})} \quad \text{and}$$
$$\frac{m_1^2}{m^2} \cdot F_p^{2/p}(\mathcal{S}) = m_1 \cdot \frac{m_1}{m^2} \cdot F_p^{2/p}(\mathcal{S}) \geq m_1.$$

Thus, conditioned on $\mathcal{B}_j^i$, $j$ is a constant $F_2$ heavy hitter in the substream $\mathcal{S}_i$. By Lemma 22, for each $j \in [t]$
$$\mathbb{P}\left[\mathcal{B}_j^i\right] \geq 1 - \frac{1}{\text{poly}(n)}$$
for another large enough poly$(n)$. Letting $\mathcal{E} = \mathcal{A} \wedge \bigwedge_{i=1}^{t} \mathcal{B}_1^i \wedge \mathcal{B}_2^i \wedge \ldots \wedge \mathcal{B}_{|H^*|}^i$, then
$$\mathbb{P}[\mathcal{E}] \geq 1 - \frac{1}{\text{poly}(n)}.$$

By the guarantee of `BPTree` and a Chernoff bound, with probability at least
$$1 - \frac{1}{\text{poly}(n)},$$
at least 0.98 fraction of the `BPTree` instances output the heavy hitters. Since the Misra-Gris algorithm is a deterministic algorithm, it outputs the heavy hitters. Lastly, for the size of $H$, by the guarantee of `BPTree`, we immediately have $|H| = \Theta(|H^*|) = \Theta(1)$. ∎

**Remark 14** *Suppose $1 \leq p < 2$ and for each $i \in \text{supp}(V(\mathcal{S}))$, $f_i \geq 1$, then*
$$\frac{\text{poly}(\epsilon^{-1}, \log n) m}{F_p^{2/p}(\mathcal{S})} \leq \frac{\text{poly}(\epsilon^{-1}, \log n) m}{F_p(\mathcal{S}) \cdot F_p^{2/p-1}(\mathcal{S})}$$
$$\leq \frac{\text{poly}(\epsilon^{-1}, \log n)}{F_p^{2/p-1}(\mathcal{S})} \ll 1,$$

as long as $F_p \gg \text{poly}(\epsilon^{-1}, \log n)$. Similarly, for $0 \leq p < 1$, we have

$$\frac{\text{poly}(\epsilon^{-1}, \log n) \, m}{F_p^{2/p}(\mathcal{S})} \leq \frac{\text{poly}(\epsilon^{-1}, \log n) \, m}{F_p^{1/p}(\mathcal{S}) \cdot F_p^{1/p}(\mathcal{S})}$$
$$\leq \frac{\text{poly}(\epsilon^{-1}, \log n)}{F_p^{1/p}(\mathcal{S})} \ll 1,$$

for sufficiently large $F_p$. Thus, the heavy hitter algorithm consumes only a small portion of the stream updates.

## 6.2 Contributions from Large Frequencies

Recall that $X_i^{(r)}$, for $r \in [k]$ and $i \in [n]$, are the limited independence $p$-inverse random variables, where $k = \Theta\left(\frac{1}{\epsilon^2}\right)$. Also recall that we wish to find the heavy hitters of the $p$-scaling transformed stream. To do so, we separate the universe items into $\Theta(\log n)$ levels, based on the values of the scaled frequencies, and thus induce $\Theta(\log n)$ substreams. A level of the scaled stream is defined from line 2 to 3 in Agorithm 3. For each $r \in [k]$, we denote the level $w$ universe items as $Z_w^{(r)}$ (formally defined in line 2). A critical level $w_0$ is chosen such that

$$2^{w_0} = \text{poly}\left(\frac{1}{\epsilon}, \log n\right),$$

and sufficiently large. In this section, we describe an algorithm that finds all the heavy hitters of the $p$-inverse scaled stream for $w \geq w_0$. For each $r \in [k]$, denote $H^{(r)}$ as follows,

$$H^{(r)} := \left\{ i \in [n] : X_i^{(r)} v_i \geq 2^{1/p-1} F_p^{1/p}(\mathcal{S}) \right\}.$$

Namely, $H^{(r)}$ is the set of heavy hitters after scaling by the $p$-Inverse random variables.

**Lemma 15** *Let $\mathcal{S}$ be a random order stream on universe $[n]$ with length $m$. Suppose $\widehat{m} = \Theta(m)$, $L \in [\text{poly}(\epsilon, \frac{1}{\log n}) F_p(\mathcal{S}), F_p(\mathcal{S})]$ and $w_0 = \Omega(\log \frac{1}{\epsilon} + \log \log n)$ is sufficiently large. Then, conditioning on two events $\mathcal{E}_1$ and $\mathcal{D}_1$, algorithm `LargeContwLen` outputs a sequence $R$ of at most $k$ numbers such that for each $r \in [k], w \geq w_0$ and $i \in H^{(r)} \cap Z_w^{(r)}$, there exists a number $y \in R$ with $y \in (1 \pm \epsilon) X_i^{(r)} v_i$. Here $\mathcal{E}_1$ depends only on the random bits of the p-ST, is independent with the order of the stream, and*

$$\mathbb{P}\left[\mathcal{E}_1\right] \geq 0.98 \quad \text{and} \quad \mathbb{P}\left[\mathcal{D}_1 | \mathcal{E}_1\right] \geq 1 - \frac{1}{\text{poly}(n)}.$$

*The algorithm uses $\mathcal{O}\left[\log n + \frac{1}{\epsilon^2}\left(\log \frac{1}{\epsilon} + \log \log n\right)\right]$ bits in the worst case and halts with at most $\mathcal{O}\left(\frac{m}{\text{poly}(\epsilon^{-1}, \log n)}\right)$ stream updates.*

**Proof** For each $i \in [n]$ and $r \in [k]$, where we recall that $k = \Theta\left(\frac{1}{\epsilon^2}\right)$, given a random order stream $\mathcal{S}$, denote

$$v := V(\mathcal{S}) := (f_1, f_2, \ldots, f_n).$$

Let $\widetilde{f}_i^{(r)} = X_i^r \cdot v_i$ be the scaled frequency of item $i$. Recall the definition of $H^{(r)}$,

$$H^{(r)} := \left\{ i \in [n] : X_i^{(r)} v_i \geq 2^{1/p-1} \|v\|_p \right\}.$$

Let $H = \bigcup_{r=1}^{k} H^{(r)}$. Let the event $\mathcal{A}$ be that $|H| = \Theta\left(\frac{1}{\epsilon^2}\right)$, which we condition on. By definition of the $p$-Inverse distribution, we obtain

$$\mathbb{P}[\mathcal{A}] \geq 0.99.$$

For each $t \in [\gamma]$ ($\gamma$ is defined in line 4), let $Z_{w,t}^{(r)} := Z_w^{(r)} \cap \{i : h(i) = t\}$. For each $i \in H$, let $w_{i,r,t}$ be the level that $i \in Z_{w_{i,r,t},t}^{(r)}$. Recall that $C = \text{poly}(\epsilon^{-1}, \log n)$ and $L \in \left[\frac{F_p^{1/p}(\mathcal{S})}{C}, F_p^{1/p}(\mathcal{S})\right]$. Hence

$$X_i^{(r)} = \Theta\left(\frac{CL}{2^{w_{i,r,t}/p}}\right).$$

Thus,

$$v_i = \Omega\left(\frac{2^{w_{i,r,t}/p} \|v\|_p}{C \, L}\right).$$

By definition of the $p$-ST, we observe, for each $i \in [n]$, $r \in [k]$ and $w \in [\Theta(\log n)]$

$$\mathbb{P}\left[i \in Z_w^{(r)} \mid \mathcal{A}\right] = \Theta\left(\frac{2^w}{C^p L^p}\right),$$

we obtain

$$\mathbb{E}\left[F_p\left(\mathcal{S}_{w,t}^{(r)}\right) \mid \mathcal{A}\right] = 2^w \cdot \frac{\|v\|_p^p}{\gamma C^p L^p} \quad \text{for any} \quad t \in [\gamma].$$

By Markov's inequality, for fixed $w, t, r$, and a fixed $i$,

$$\mathbb{P}\left[F_p\left(\mathcal{S}_{w,t}^{(r)} \setminus \{i\}\right) \geq \mathcal{O}\left(\frac{2^w}{C^p} \frac{\|v\|_p^p}{L^p}\right) \Big| \mathcal{A}\right] \leq \frac{1}{\gamma} \leq \mathcal{O}(\epsilon^2).$$

Conditioned on $\mathcal{A}$, $|H| = \mathcal{O}(k) = \mathcal{O}\left(\frac{1}{\epsilon^2}\right)$. Then, by a union bound, with probability at least $0.99$, for every $r \in [k]$, each $i \in H^{(r)}$ is a $(0.5, F_p)$-heavy hitter in a substream $\mathcal{S}_{w,t}^{(r)}$ for some $w, t$ (as long as $\gamma = \Omega(\epsilon^{-2})$ is large enough). Let this event be $\mathcal{E}_1$. Note that event $\mathcal{E}_1$ is independent with the order of the stream.

Now consider the su-routine of estimating the length of the stream $\mathcal{S}_{wt}^{(r)}$ in line 13 of Algorithm 3. If $i \in Z_{w,t}^{(r)}$, then

$$F_1\left(\mathcal{S}_{w,t}^{(r)}\right) \geq f_i(\mathcal{S}) \gtrsim \frac{2^{w/p} \|v\|_p}{C \, L} = \Omega\left[\text{poly}\left(\frac{1}{\epsilon}, \log n\right)\right].$$

Denote $m_1 = m_1' \widehat{m}/m$, where

$$m_1' = \frac{m}{z_w} F_1\left(\mathcal{S}_{w,t}^{(r) t_c:(t_c+z_w)}\right).$$

Therefore, by Lemma 22 (with the same argument as in the proof of Proposition 18), with probability at least $1 - 1/\operatorname{poly}(n)$, the estimated length $m'_1$ satisfies,

$$\left|m'_1 - F_1(\mathcal{S}^{(r)}_{w,t})\right| \leq 4\sqrt{\tfrac{m \cdot \log n}{z_w \cdot F_1(\mathcal{S}^{(r)}_{w,t})}} F_1(\mathcal{S}^{(r)}_{w,t}) = \mathcal{O}(F_1(\mathcal{S}^{(r)}_{w,t}))$$

for sufficiently large $z_w$. Thus $m_1$ is a constant approximation to $F_1(\mathcal{S}^{(r)}_{w,t})$. Thus, in line 15, with probability at least $1 - 1/\operatorname{poly}(n)$, every element $i \in H^{(r)} \cap Z^{(r)}_{w,t}$ is in the output of the HHR algorithm. Similarly, with probability at least $1 - 1/\operatorname{poly}(n)$, line 16 returns a $(1 \pm \epsilon)$ approximation to $f_i(\mathcal{S})$ for each $i \in H^{(r)} \cap Z^{(r)}_{w,t}$.

For those sub-streams containing no heavy hitters, the set output by HHR is arbitrary. For each such item $j$, suppose it is output in level $w$. Then $v_j = O\left(\frac{2^{w/p}\|v\|_p}{CL}\right)$. By Proposition 18, with probability at least $1 - 1/\operatorname{poly}(n)$, the estimated frequency of each of them does not exceeds $v_j + \epsilon 2^{w/p}\|v\|_p/(LC)$. Thus $j$ would not be stored by the heap structure. Let $\mathcal{D}_1$ be the event that the above happens for all $i, r$. We obtain that

$$\mathbb{P}\left[\mathcal{D}_1 | \mathcal{E}_1\right] \geq 1 - \frac{1}{\operatorname{poly}(n)},$$

for sufficiently large $w_0$.

The last step in line 18 stores a $(1 \pm \epsilon)$ approximation of the estimated frequency in a heap data structure. This data structure uses at most

$$\mathcal{O}\left[\frac{1}{\epsilon^2}\left(\log \frac{1}{\epsilon} + \log \log n\right)\right]$$

bits.

It remains to show that the number of stream updates consumed by the algorithm is bounded. Since $w_0 = \Omega(\log \frac{1}{\epsilon} + \log \log n)$ can be made sufficiently large, the algorithm halts after seeing at most

$$\frac{m}{\operatorname{poly}\left(\epsilon^{-1}, \log n\right)}$$

stream updates. ∎

**Theorem 16** *Let $\mathcal{S}$ be a random order stream on universe $[n]$ with length $m$. Suppose $\operatorname{poly}(\epsilon, 1/\log n) F_p(\mathcal{S}) \leq L \leq F_p(\mathcal{S})$ and $w_0 = \Omega(\log \frac{1}{\epsilon} + \log \log n)$. Then Algorithm 4 has the same guarantee as stated in Lemma 15.*

**Proof** Algorithm 4 introduces a deterministic guess of the length of the stream. The correctness thus follows directly from Lemma 15. ∎

**Algorithm 3:** Estimate the contribution of large frequencies: `LargeContwLen`$(p, n, \epsilon, w_0, L, \widehat{m}, \{X_i^r\}, \mathcal{S})$

**Data:**
$L \in \left[\frac{F_p^{1/p}(\mathcal{S})}{C}, F_p^{1/p}(\mathcal{S})\right]$, where $C = \text{poly}(\epsilon^{-1}, \log n)$;
$\mathcal{S} = \langle a_1, a_2, \ldots, a_m \rangle$ is random order stream of length $m$;
$w_0$, the smallest level to be considered;
$\widehat{m} \in (\frac{m}{c_1}, c_1 m)$ is a constant approximation to $m$;
$\{X_i^r\}$ for $i = 1, 2, \ldots, n$, $r = 1, 2, \ldots, k$ are pairwise independent $p$-Inverse random variables;

**Result:** $R = (r_1, r_2, \ldots, r_k)$, where each $r_i \in \mathbb{R}$;

1 **Initialize:**
2    $Z_0^{(r)} \leftarrow \left\{i \in [n] \times [k] : X_i^{(r)} \geq CL\right\}$; $Z_w^{(r)} \leftarrow \left\{i \in [n] \times [k] : \frac{CL}{2^{(l-1)/p}} < X_i^{(r)} \leq \frac{CL}{2^{w/p}}\right\}$;
3    $\mathcal{S}_w^{(r)} \leftarrow S|_{Z_w^{(r)}}$ the substream induced by $Z_w^{(r)}$, and $\mathcal{S}_w^{(r) m_1:m_2} \leftarrow S^{m_1:m_2}|_{Z_w^{(r)}}$;
4    $\gamma \leftarrow \Theta\left(\frac{1}{\epsilon^2}\right)$, $h : [n] \to [\gamma]$, pairwise independent hash function;
5    HEAP($k$): a heap structure to maintain the largest-$k$ inputs;
6    $t_c \leftarrow 1$: the current pointer to the stream updates;
7    $z_w \leftarrow \frac{\widehat{m} C \text{poly}(\epsilon^{-1}, \log n)}{2^{w/p}}$;   /*$z_w \ll \widehat{m}$*/;
8 **if** *Stream ends before output* **then**
9    **return** Fail;
10 **for** $w = w_0 + 1, w_0 + 2, \ldots, \Theta(\log n)$ **do**
11    **for** *each* $t \in [\gamma], r \in [k]$ **do**
12       $\mathcal{S}_{wt}^{(r)} \leftarrow$ substream induced by $Z_w^{(r)} \cap \{i \in [n] : h(i) = t\}$;
13       $m_1 \leftarrow$ estimate the length of $\mathcal{S}_{wt}^{(r)}$ using $\mathcal{S}^{t_c:(t_c+z_w)}$ updates; $t_c \leftarrow t_c + z_w$ ;
14       $N \leftarrow$ HHR$\left(p, c_0, m_1, \mathcal{O}\left(\frac{2^w}{C^p}\right), \mathcal{S}_{wt}^{(r) t_c:-1}\right)$.output(), for some small enough constant $c_0 > 0$;   /*$c_0 \approx 0.5$*/;
15       $t_c \leftarrow t_c +$ stream length consumed by HHR;
16       $R_f \leftarrow$ QueryFrequency$\left(2^{w_0}, N, \epsilon, \mathcal{S}^{t_c:(t_c+z_w)}\right)$; $t_c \leftarrow t_c + z_w$;
17       **for** $a \in N$ **do**
18          HEAP.update $\left\lceil \frac{\log\left(X_a^{(r)} \cdot R_f(a)\right)}{\log(1+\epsilon)} \right\rceil$;

19 **return** HEAP.output();

## 6.3 Contributions from Small Frequencies

In this section, we describe the algorithm that approximates all heavy scaled frequencies in small levels, i.e., $w \leq w_0$. In those levels, $X_i^{(r)}$'s are large. Therefore, the probability of sampling such a large $X_i^{(r)}$ is small. Thus the resulting sets $Z_w^{(r)}$ are small enough, i.e., of size $\text{poly}(\epsilon^{-1}, \log n)$, with high probability. If the stream length of $\mathcal{S}_w^{(r)}$ is also small (e.g., $\text{poly}(\epsilon^{-1}, \log n)$), we can then directly run BPTree on these streams to report all the IDs (i.e.,

---

**Algorithm 4:** Estimating the contribution of large frequencies: `LargeCont`$(p, n, \epsilon, w_0, L, \{X_i^r\}, \mathcal{S})$

**Result:** $R = (r_1, r_2, \ldots, r_k)$, where each $r_i \in \mathbb{R}$;

1. **Initialize:**
2.     $\widehat{m} \leftarrow 2$;
3.     $m_1 \leftarrow 0$;
4.     $A_1 \leftarrow$ `LargeContwLen`$(p, n, \epsilon, w_0, L, \widehat{m}, \{X_i^r\}, \mathcal{S})$;
5.     $A_2 \leftarrow$ `LargeContwLen`$(p, n, \epsilon, w_0, L, \widehat{m}, \{X_i^r\}, \mathcal{S})$;
6. **while** *Not the end of the stream* **do**
7.     Read an input $a$;
8.     Input $a$ to $A_1$ and $A_2$;
9.     $m \leftarrow m + 1$;
10.     Input $a$ to `LargeContwLen`;
11.     **if** $2\widehat{m} \leq m_1$ **then**
12.         Discard $A_1$;
13.         $A_1 \leftarrow A_2$;
14.         $\widehat{m} \leftarrow m_1$;
15.         $A_2 \leftarrow$
16.         `LargeContwLen`$(p, n, \epsilon, w_0, L, \widehat{m}, \{X_i^r\}, \mathcal{S}^{m:-1})$;
17.     **return** whatever $A_1$ returns;

---

hashed IDs) of the heavy hitters. We then use a modified `CountSketch` instance to report the frequencies of these heavy hitters.

We modify the `CountSketch` [CCFC02] algorithm as follows. Each counter of the `CountSketch` is restricted to be no more than $\mathcal{O}(\log\log n + \log 1/\epsilon)$ bits. If a counter exceeds this number of bits, we put an $\infty$ marker on the counter. It is easy to see that if $f_i = \mathcal{O}(\text{poly}(\log n, \frac{1}{\epsilon}))$, then the $i'$-frequency can be reported as in the original algorithm. However if $f_i = \omega(\text{poly}(\log n, 1/\epsilon))$, then one can simply report $\infty$. If we have that the universe size is at most $\text{poly}(\log n, 1/\epsilon)$, then the modified algorithm will use $\mathcal{O}\left[\frac{1}{\epsilon^2} \log \frac{1}{\epsilon\delta}\big((\log\log n + \log 1/\epsilon)\big)\right]$ bits of space for error probability at most $0 < \delta < 1$.

However, if the stream length is large, i.e., $\text{poly}(n)$, we can simply pick the initial $\Theta(\text{poly}(\epsilon^{-1}, \log n))$-length portion of the stream to estimate the heavy hitters and the frequencies. This is because in a random order stream, one can use a portion of the stream to sample sufficiently many stream updates of the heavy hitters. Hence we obtain a good approximation. Since we can consider only a small universe and a small stream length, `BPTree` works with the desired space. The guarantee of the algorithm is as follows.

**Theorem 17** *Let $\mathcal{S}$ be a random order stream on universe $[n]$ with length $m$. Then, conditioned on three events $\mathcal{E}_2, \mathcal{F}_2$ and $\mathcal{D}_2$, algorithm $\mathtt{SmallCont}$ outputs a sequence $R$ of at most $k$ numbers such that for each $r \in [k], w < w_0$ and $i \in H^{(r)} \cap Z_w^{(r)}$, there exists a number $y \in R$ with $y \in (1 \pm \epsilon) X_i^{(r)} v_i$. Here $\mathcal{E}_2$ depends only on the random bits of the p-ST and is independent with the order of the stream. $\mathcal{F}_2$ is also independent of the stream conditioning* on $\mathcal{D}_2$ and $\mathcal{E}_2$. Furthermore,

$$\mathbb{P}\left[\mathcal{E}_2\right] \geq 0.98, \quad \mathbb{P}\left[\mathcal{D}_2 | \mathcal{E}_2\right] \geq 1 - \frac{1}{\text{poly}(n)} \quad \text{and}$$

$$\mathbb{P}\left[\mathcal{F}_2 | \mathcal{E}_2, \mathcal{D}_2\right] \geq 0.9.$$

The algorithm uses $\mathcal{O}[\log n + \frac{1}{\epsilon^2}\left(\log \frac{1}{\epsilon} + \log\log n\right)^4]$ bits in the worst case.

**Proof** By the same argument of the proof of Lemma 15, with probability at least 0.98, if for each $i \in H^{(r)} \cap Z_w^{(r)}$ for all $w \leq w_0$, $i$ is a $\left(\Omega\left(\frac{1}{w_0}\right), F_p\right)$-heavy hitter of the stream $\mathcal{S}_w^{(r)}$. Let this event be $\mathcal{E}_2$, which only depends on the random bits of the $p$-ST transformation. For the rest of the proof, we condition on this event.

For each $w \in [w_0]$, let $m_w$ be the length of the stream BPTree used to approximate the heavy hitter. Let event $\mathcal{D}_2$ be that for each $r \in [k], w \leq w_0$ and $i \in H^{(r)} \cap Z_w^{(r)}$, $i$ is a $\left(\Theta\left(\frac{1}{w_0}\right), F_2\right)$-heavy hitter in $\mathcal{S}_w^{(r),:m_w}$ and

$$\frac{f_i\left(\mathcal{S}_w^{(r),:m_w}\right) m}{m_w} \in (1 \pm \epsilon) f_i(\mathcal{S}).$$

We argue that $\mathbb{P}(\mathcal{D}_2) \geq 1 - \frac{1}{\text{poly}(n)}$. First, if $F_p\left(\mathcal{S}_w^{(r)}\right) = \mathcal{O}(\text{poly}(\epsilon^{-1}, \log n))$, then by definition of $m_w$, we have that $m_w = m$ and the claim is true for any $i \in H^{(r)} \cap Z_w^{(r)}$. Next, we consider $F_p\left(\mathcal{S}_w^{(r)}\right) = \Omega(\text{poly}(\epsilon^{-1}, \log n))$ and sufficiently large, i.e., the memory exceeds in Line 12. We have that $m_w = \Theta\left(\text{poly}(\epsilon^{-1}, \log n)\right)$. By the same argument as in the proof of Theorem 13, i.e., picking

$$m_w = \Omega\left[\frac{F_1\left(\mathcal{S}_w^{(r)}\right)^2}{F_p^{2/p}\left(\mathcal{S}_w^{(r)}\right)}\right],$$

then with probability at least $1 - 1/\text{poly}(n)$, any $i \in H^{(r)} \cap Z_w^{(r)}$ is a $(\Theta(1), F_2)$-heavy hitter in stream $\mathcal{S}_w^{(r)}$. By Lemma 24, with probability at least $1 - 1/\text{poly}(n)$, for each $w \leq w_0$, $m_w = \text{poly}(\epsilon^{-1}, \log n)$.

Lastly, let event $\mathcal{F}_2$ be the event that all the HHR instances are correct and all the CountSketch instances are correct. It follows immediately that $\mathbb{P}(\mathcal{F}_2) \geq 0.9$. Conditioned on $\mathcal{F}_2$, it is easy to see that the frequencies of each heavy hitter are reported with the desired precision.

It remains to bound the space used by the algorithm. Each BPTree instance takes $\mathcal{O}\big(w_0 \log w_0 \cdot (\log \log n + \log \frac{1}{\epsilon})\big)$ bits of space. Each modified count sketch instance takes

$$\mathcal{O}\left[\frac{w_0}{\epsilon^2} \log \frac{w_0}{\epsilon} \left(\log \log n + \log \frac{1}{\epsilon}\right)\right]$$

bits of space. There are in total $\mathcal{O}(w_0)$ count sketch instances and $\mathcal{O}(w_0/\epsilon^2)$ BPTree instances. Therefore, the total space used is $\mathcal{O}\big[\frac{1}{\epsilon^2}\left(\log\log n + \log \frac{1}{\epsilon}\right)^3 \cdot \left(\log\log\log n + \log \frac{1}{\epsilon}\right)\big]$. ∎

## 6.4 Frequency Query in the Random Order Model

If an item has large frequency, we can estimate its frequency with a small number of stream updates. The formal algorithm is presented in Algorithm 6. It has the following guarantee:

**Proposition 18** *Given a random order stream $\mathcal{S}$ on universe $[n]$ with length $m$. For a fixed $i^*$, if $\frac{\widehat{m} f_{i^*}(\mathcal{S})}{m} = \Omega\left(\frac{1}{\epsilon^2} \log \frac{1}{\delta}\right)$ then, with probability at least $1 - \delta$, Algorithm 6 returns a number $f$, s.t. $\frac{mf}{\widehat{m}} \in (1 \pm \epsilon) f_i(\mathcal{S})$. Otherwise, $\frac{mf}{\widehat{m}} = \mathcal{O}\left(\frac{m \log \frac{1}{\delta}}{\widehat{m}}\right)$. The algorithm uses $\mathcal{O}(\log n)$ bits of space and consumes $\widehat{m}$ stream updates.*

**Proof** By Lemma 22, with probability at least $1 - \delta$,

$$\left| \frac{f_{i^*}\left(\mathcal{S}^{:\widehat{m}}\right)}{\widehat{m}} - \frac{f_{i^*}(\mathcal{S})}{m} \right| \leq 4 \sqrt{\frac{\log \frac{1}{\delta}}{\widehat{m}}} \max\left( \sqrt{\frac{f_{i^*}(\mathcal{S})}{m}}, \sqrt{\frac{\log \frac{1}{\delta}}{\widehat{m}}} \right),$$

which we condition on. Consider Case 1: $\frac{\widehat{m} f_{i^*}(\mathcal{S})}{m} = \Omega\left(\frac{1}{\epsilon^2} \log \frac{1}{\delta}\right)$, we obtain,

$$\left| \frac{m f_{i^*}\left(\mathcal{S}^{:\widehat{m}}\right)}{\widehat{m}} - f_{i^*}(\mathcal{S}) \right| \leq \epsilon f_{i^*}(\mathcal{S}).$$

Now consider case 2: $\frac{\widehat{m} f_{i^*}(\mathcal{S})}{m} = \mathcal{O}\left(\frac{1}{\epsilon^2} \log \frac{1}{\delta}\right)$. We obtain,

$$\left| \frac{m f_{i^*}\left(\mathcal{S}^{:\widehat{m}}\right)}{\widehat{m}} - f_{i^*}(\mathcal{S}) \right| = \mathcal{O}\left(\frac{m \log \frac{1}{\delta}}{\widehat{m}}\right).$$

∎

# 7 Deterministic Algorithm for $F_p$ Approximation

In this section we introduce our deterministic algorithm for $F_p$ approximation, which follows from our randomized algorithm with an initial space-efficient randomness extraction procedure applied to a prefix of the stream.

## 7.1 Upper Bound

**Theorem 19** *Fix $p \geq 0, \epsilon \in (0, 1)$. There exists a deterministic algorithm that makes a single pass over a random order stream $\mathcal{S}$ on the universe $[n]$, and outputs a number $F \in (1 \pm \epsilon) F_p(\mathcal{S})$ with probability at least $1 - \delta$, where the randomness is over the order of the stream updates. The algorithm uses*

$$\mathcal{O}\Big[\frac{1}{\epsilon^2} \left(\log \log n + \log \frac{1}{\epsilon}\right)^4 \log \frac{1}{\delta} + \log n \cdot (\log \log n + \log \frac{1}{\delta}) \cdot \frac{1}{\delta} + \log n \log \frac{1}{\epsilon}\Big]$$

*bits of memory, provided $\delta \geq 1/\operatorname{poly}(n)$.*

**Proof** W.l.o.g., we assume $\epsilon \geq 1/\sqrt{n}$, since otherwise we can simply store an approximate counter for each item in the stream. It is sufficient to show that we are able to de-randomize the randomized algorithm using the random updates from the stream using a near-logarithmic number of bits of space. First we pick $s = \mathcal{O}(\log \log n + \log \delta^{-1})$ and store all the universe items, their frequencies and their first arrival times until we obtain $s$ distinct items in the stream. Let $z_1$ denote when the this happens. We assume the stream is long enough that this step can be done, since otherwise we obtain an exact estimate of $F_p$. We show in Section 7.1.1 how to obtain a nearly uniformly random seed of $s$ bits.

We thus obtain a nearly uniform sample from all the prime numbers with $\mathcal{O}(\log \log n + \log \frac{1}{\delta})$ bits (note that there are $\mathrm{poly}(\log n/\delta)$ many such prime numbers). Let this sampled prime number be $q$. For the next $\mathcal{O}(\log n/\delta)$ distinct universe items, denoted by $R$, we argue that with probability at least $1 - \delta$, all of them are distinct modulo $q$. Indeed, consider any $r_1, r_2 \in R$ with $r_1 \neq r_2$. Then $r_1 - r_2$ can have at most $\log n$ prime factors ([HW79], p.355). For all $\binom{|R|}{2}$ pairs, their differences can have at most $\mathcal{O}(\log^3 n/\delta^3)$ distinct prime factors in total. Thus with probability at least $1 - \delta$, $q$ does not divide any of the differences. Therefore the set $R$ is mapped to distinct numbers modulo $q$. The value $q$ can be stored using at $\mathcal{O}(\log \log n + \log \frac{1}{\delta})$ bits.

Next we approximately store the frequencies of each item in $R$ using the random order stream. To do this, we first fix the following numbers $g_1 = 1, g_2 = 2, g_3 = 4, \ldots, g_i = 2^{i-1}$. For each $r \in R$, we store the largest number $i$ such that $f_r(\mathcal{S}^{z_1:g_i}) = \mathrm{poly}(\log n, \epsilon^{-1})$ and $f_r(\mathcal{S}^{z_1:g_i})$ as well. Therefore, such an operation only costs $\mathcal{O}(\frac{\log n}{\delta}(\log \log n + \log \frac{1}{\delta} + \log \epsilon^{-1}))$ bits. By Lemma 23, the frequency of each item is preserved up to a $(1 \pm \epsilon)$ factor with probability at least $1 - 1/\mathrm{poly}(n)$. Note that if the stream ends before we observe all of $R$, we obtain a good approximation to $F_p(\mathcal{S})$ immediately. We also store the first occurrence in the stream of each item in $R$. We also store the parity of the first appearance time of each item.

Repeating the extraction argument in Section 7.1.1 for the set $R$, we can now extract $\mathcal{O}(\log n)$ bits that is $(1 \pm \delta)$ close to uniform. Given these bits, it now suffices to run our earlier randomized algorithm on the remaining part of the stream.

There is one last problem remaining, however. Namely, it may be the case that the stream used for extracting random bits contributes too much to $F_p(\mathcal{S})$, causing the estimation of the randomized algorithm to have too much error (since the prefix and the suffix of the stream share the same items, we need to take the $p$-th power of the sum of their frequencies). This problem can be solved as follows – we can continue the frequency estimation in parallel with the randomized algorithm until the $F_p$ value becomes at least a $1/\epsilon$ factor larger than the time when we initialized our randomized algorithm. Therefore, if the stream ends before this happens, then we use the frequency estimates for calculating $F_p(\mathcal{S})$ from our deterministic algorithm. Otherwise we use the value of the randomized algorithm (which is seeded with the seed found by our deterministic algorithm). In either case, the overall error is at most $\epsilon F_p$. ∎

### 7.1.1 Derandomization

Let $s > 0$ be a parameter. Suppose we store $t = \mathcal{O}\left(\frac{s}{\delta}\right)$ distinct universe items with their approximate frequencies as well as the IDs and the parities of their first appearances in the stream. Denote the set of these items as $H$ and the overall length of the stream as $m'$. Firstly, we sort the items by their approximate counts and take the smallest $\delta t/100$ items as set $L$. We additionally sort the items in $L$ by their IDs, and obtain a bits string $b$ of length $|L|$, where each bit $b_i$ is the parity of the first appearance of the $i$-th item in $L$. Since $L$ contains the smallest $\delta t/100$ items of $H$, we have for each $w \in L$, $f_w(\mathcal{S}^{0:m'}) \leq m'/t/(1-\delta/100)$. Thus for each bit $b_i$,

$$\mathbb{P}[b_i = 0], \mathbb{P}[b_i = 1] \in \frac{1}{2} \pm \frac{2}{m'}.$$

and

$$\forall x \in \{0,1\}^{|L|} : \mathbb{P}[b = x] \in \left(\frac{1}{2} \pm \frac{2}{m'}\right)^{|L|} \subset \frac{1}{2^{|L|}}\left(1 \pm \frac{5|L|}{m'}\right) \subset \frac{1}{2^{|L|}}\left(1 \pm \frac{\delta}{20}\right).$$

As such, we obtain a bits stream of length $\Omega(s)$, that is close to uniform bits up to a $(1 \pm \delta)$ factor.

## A  Proofs

**Lemma 20**
$$(1-x)^{-\alpha} \geq 1 + \alpha x \quad \text{and} \quad (1-x)^{-\alpha} \leq 2$$
for any $x \in \left(0, \frac{1}{2\alpha}\right)$ and $\alpha \geq 0$.

**Lemma 21**
$$(1+x)^{-\alpha} \leq 1 - \frac{\alpha}{2}x \quad \text{and} \quad (1+x)^{-\alpha} \leq 1$$
for any $x \in \left(0, \frac{1}{2\alpha}\right)$ and $\alpha \geq 0$.

**Proof** [ of Lemma 4] For each $(i,j) \in [k] \times [n]$, $f_{(i,j)}(\mathcal{S}') = f_j(\mathcal{S}) X_{i,j}$, where $X_{i,j}$ is an $\alpha$-Inverse random variable. We define the indicator random variables

$$A_{i,j} := \mathbb{I}\left\{f_{(i,j)}(\mathcal{S}') \geq 2^{1/\alpha}(1-\epsilon)\|V(\mathcal{S})\|_\alpha\right\} \quad \text{and}$$
$$B_{i,j} := \mathbb{I}\left\{f_{(i,j)}(\mathcal{S}') \geq 2^{1/\alpha}(1+\epsilon)\|V(\mathcal{S})\|_\alpha\right\}.$$

Thus
$$|U_+| = \sum_{(i,j)\in[k]\times[n]} A_{i,j} \quad \text{and} \quad |\overline{U}_-| = \sum_{(i,j)\in[k]\times[n]} B_{i,j}.$$

By definition of the $\alpha$-Inverse distribution,

$$\mathbb{P}\left[A_{i,j} = 1\right] = \frac{1}{2}\left(1 - \epsilon\right)^{-\alpha} \frac{f_{i,j}(\mathcal{S}')^\alpha}{\|V(\mathcal{S})\|_\alpha^\alpha} \quad \text{and}$$

$$\mathbb{P}\left[B_{i,j} = 1\right] = \frac{1}{2}\left(1 + \epsilon\right)^{-\alpha} \frac{f_{i,j}(\mathcal{S}')^\alpha}{\|V(\mathcal{S})\|_\alpha^\alpha}.$$

Hence,

$$\mathbb{E}\left(|U_+|\right) = \frac{k}{2}\left(1 - \epsilon\right)^{-\alpha} \geq \frac{k}{2}\left(1 + \alpha\epsilon\right) \quad \text{and}$$

$$\mathbb{V}(|U_+|) \leq \frac{k}{2}\left(1 - \epsilon\right)^{-\alpha} \leq k.$$

By Chebishev's inequality,

$$\mathbb{P}\left[\sum_{i \in [k], j \in [n]} A_{i,j} < \frac{1}{2}k\right] \leq \frac{k}{\left(\frac{\alpha}{2}\epsilon k\right)^2} = \left(\frac{4}{\alpha^2}\right)\frac{1}{\epsilon^2 k} \leq 0.05,$$

for $k \geq 160/(\alpha^2\epsilon^2)$. Thus,

$$\mathbb{P}\left[|U_+| < \frac{1}{2}k\right] \leq 0.05.$$

Similarly,

$$\mathbb{P}\left[|\overline{U}_-| \geq \frac{1}{2}k\right] \leq 0.05.$$

By a union bound, with probability at least 0.1, we have

$$|U|_+ \geq \frac{k}{2} \quad \text{and} \quad |\overline{U}_-| < \frac{k}{2},$$

which we condition on. Thus

$$|U_+ \cap U_-| = |U_+ \backslash \overline{U}_-| > \frac{k}{2} - \frac{k}{2} = 0,$$

concluding the proof. ∎

**Lemma 22** *Let $\mathcal{S} = \langle a_1, a_2, \ldots, a_m \rangle$ be a random order stream on the universe $[n]$. Let $\mathcal{S}' = \langle a_{i_1}, a_{i_2}, \ldots, a_{i_k} \rangle$ be a substream of fixed $k$ distinct integers $\{i_1, i_2, \ldots, i_k\}$ with $k \leq m$. Then, for any $j \in [n]$*

$$\mathbb{P}\left[\left|\frac{f_j(\mathcal{S}')}{k} - \frac{f_j(\mathcal{S})}{m}\right| \geq 4\sqrt{\frac{\log\frac{1}{\delta}}{k}} \max\left(\sqrt{\frac{f_j(\mathcal{S})}{m}}, \sqrt{\frac{\log\frac{1}{\delta}}{k}}\right)\right]$$
$$\leq 2\delta.$$

**Proof** We assign unique labels to each of the updates of item $j$, i.e., let these updates be

$$b_1, b_2, \ldots, b_{f_j(\mathcal{S})}.$$

Let

$$Z_1, Z_2, \ldots, Z_{f_j(\mathcal{S})}$$

be the indicators of these items to be in $\mathcal{S}'$. Then

$$\mathbb{P}\left[Z_i = 1\right] = \frac{\binom{m-1}{k-1}}{\binom{m}{k}} = \frac{(m-1)!}{(k-1)!(m-k)!} \cdot \frac{k!(m-k)!}{m!} = \frac{k}{m}.$$

We now consider the Doob Martingale $\mathbb{E}[\sum_{l=1}^{f_j} Z_l | Z_1, Z_2, \ldots, Z_r]$ for $r = 1, 2, \ldots, f_j$. Let

$$D_r = \mathbb{E}\left[\sum_{l=1}^{r} Z_l | Z_1, Z_2, \ldots, Z_r\right] - \mathbb{E}\left[\sum_{l=1}^{r} Z_l | Z_1, Z_2, \ldots, Z_{r-1}\right].$$

We have

$$\mathbb{P}\left[D_r = 1\right] \leq \frac{k - \sum_{l=1}^{j-1} Z_l}{m - \sum_{l=1}^{j-1} Z_l} \leq \frac{k}{m}.$$

Thus

$$\mathbb{V}\left(D_r | Z_1, Z_2, \ldots, Z_{r-1}\right) \leq \frac{k}{m}.$$

By Bernstein's inequality [DP09](Theorem 8.2),

$$\mathbb{P}\left[\left|\sum_l Z_l - \frac{k}{m} \cdot f_j(\mathcal{S})\right| > \gamma\right] \leq 2 \exp\left[\frac{-\gamma^2}{2V[1 + \frac{\gamma}{3V}]}\right],$$

where

$$V = \sum_{r=1}^{f_j} \mathbb{V}\left(D_r | Z_1, Z_2, \ldots, Z_{r-1}\right) \leq \frac{k}{m} \cdot f_j(\mathcal{S}).$$

Plugging in $\gamma = 4\sqrt{\log \frac{1}{\delta}} \max\left(\sqrt{\frac{k}{m} f_j(\mathcal{S})}, \sqrt{\log \frac{1}{\delta}}\right)$, we obtain

$$\mathbb{P}\left[\left|\sum_l Z_l - \frac{k}{m} \cdot f_j(\mathcal{S})\right| > \gamma\right] \leq 2 \exp\left[-\log \frac{1}{\delta}\right] \leq 2\delta.$$

∎

**Lemma 23** Let $\mathcal{S} = \langle a_1, a_2, \ldots, a_m \rangle$ be a random order stream on the universe $[n]$. Let $\mathcal{S}' = \langle a_{i_1}, a_{i_2}, \ldots, a_{i_k} \rangle$ be a substream of $k$ fixed distinct integers $\{i_1, i_2, \ldots, i_k\}$ with $k \leq m$. Then,

$$\mathbb{P}\left[\left|\frac{F_1(\mathcal{S}')}{k} - \frac{F_1(\mathcal{S})}{m}\right| \geq 4\sqrt{\frac{\log \frac{1}{\delta}}{k}} \max\left(\sqrt{\frac{F_1(\mathcal{S})}{m}}, \sqrt{\frac{\log \frac{1}{\delta}}{k}}\right)\right] \leq 2\delta.$$

**Proof** This is identical to the proof of Lemma 22. ∎

**Lemma 24** Let $\mathcal{S} = \langle a_1, a_2, \ldots, a_m \rangle$ be a random order stream on the universe $[n]$. Let $\mathcal{S}' = \langle a_{i_1}, a_{i_2}, \ldots, a_{i_k} \rangle$ be a substream of $k$ fixed distinct integers $\{i_1, i_2, \ldots, i_k\}$ with $k \leq m$. Suppose $F_0(\mathcal{S}) \geq \epsilon^{-2}$ and for each $i \in [n]$, $k \leq m - f_i$. Then,

$$\sum_{i \in [n]: f_i > 0} \left[1 - \left(1 - \frac{k}{m - f_i}\right)^{f_i}\right] \leq \mathbb{E}(F_0(\mathcal{S}'))$$

$$\leq \sum_{i \in [n]: f_i > 0} \left[1 - \left(1 - \frac{k}{m}\right)^{f_i}\right]$$

and

$$\mathbb{P}[|F_0(\mathcal{S}') - \mathbb{E}(F_0(\mathcal{S}'))| \geq \epsilon F_0(\mathcal{S})] \leq \exp\left(-\epsilon^2 F_0(\mathcal{S})\right).$$

**Proof** Consider stream $\mathcal{S}'$, and let $Z_i$ denote the indicator random variable indicating that element $i$ is sampled. We have that

$$\mathbb{P}[Z_i = 1] = 1 - \frac{\binom{m - f_i}{k}}{\binom{m}{k}}$$

$$= \frac{(m - f_i - k + 1)(m - f_i - k + 2) \ldots (m - k)}{(m - f_i + 1)(m - f_i + 2) \ldots (m)}.$$

Thus

$$\left(1 - \frac{k}{m - f_i}\right)^{f_i} \leq \mathbb{P}[Z_i = 1] \leq \left(1 - \frac{k}{m}\right)^{f_i}.$$

For concentration, we will use a Bernstein inequality for martingales [DP09](Theorem 5.1). For each $Z_i$, when it is revealed, $F_0$ changes by at most 1. Thus,

$$\mathbb{P}[|F_0(\mathcal{S}') - \mathbb{E}(F_0(\mathcal{S}'))| > \epsilon F_0(\mathcal{S})] \leq 2\exp\left[-\epsilon^2 F_0(\mathcal{S})\right].$$

∎

**Lemma 25 (Substream polylog Approx.)** *Let $\mathcal{S} = \langle a_1, a_2, \ldots, a_m \rangle$ be a random order stream on the universe $[n]$. Let $\mathcal{S}' = \langle a_{i_1}, a_{i_2}, \ldots, a_{i_k} \rangle$ be a substream of $k$ fixed distinct integers $K = \{i_1, i_2, \ldots, i_k\}$ with $k \leq m$. Then for fixed $p \geq 0, \delta \in (0, 1/2)$, with probability at least $1 - \delta$,*

$$\min\left(1, \frac{5m^2}{k^2}\left(\log \frac{20n}{\delta}\right)^{1-p}\right) \frac{k^p}{(5m)^p} F_p(\mathcal{S}) - 4\log \frac{20n}{\delta} \leq F_p(\mathcal{S}').$$

**Proof** Given $\delta' \in \left(0, \frac{\delta}{20n}\right]$, we split the universe $[n]$ into two sets,

$$\mathcal{N}_1 = \left\{i \in [n] : f_i \geq \frac{5m\log\frac{1}{\delta'}}{k}\right\} \quad \text{and} \quad \mathcal{N}_2 = [n]\setminus\mathcal{N}_1.$$

Accordingly, define

$$\mathcal{S}_1 = \langle a_j : a_j \in \mathcal{N}_1 \rangle \quad \text{and} \quad \mathcal{S}_2 = \langle a_j : a_j \in \mathcal{N}_2 \rangle.$$

$\mathcal{S}'_1$ and $\mathcal{S}'_2$ are defined similarly for the set of $K$.

First, by Lemma 22, with probability at least $1 - 2n\delta'$, for each $i \in \mathcal{N}_1$,

$$f_i(\mathcal{S}') = \frac{k}{m}f_i(\mathcal{S}) \pm 4\sqrt{\log \frac{1}{\delta'}}\sqrt{\frac{kf_j(\mathcal{S})}{m}}$$

$$\Rightarrow \frac{k}{5m}f_i(\mathcal{S}) \leq f_i(\mathcal{S}') \leq \frac{3k}{m}f_i(\mathcal{S}).$$

For the remainder of $\mathcal{N}_2$, with probability at least $1 - 2\delta'$,

$$F_1(\mathcal{S}'_2) \geq \max\left\{\frac{k}{m}F_1(\mathcal{S}_2) - 4\log\frac{1}{\delta'}, 0\right\}.$$

Therefore, with probability at least $1 - (2n+2)\delta'$,

$$F_p(\mathcal{S}') \geq \frac{k^p}{5^p m^p}F_p(\mathcal{S}_1) + \max\left\{\frac{k}{m}F_1(\mathcal{S}_2) - 4\log\frac{1}{\delta'}, 0\right\}$$

$$\geq \frac{k^p}{5^p m^p}F_p(\mathcal{S}_1) + \left(\frac{5m\log\frac{1}{\delta'}}{k}\right)^{1-p} \frac{k}{m}F_p(\mathcal{S}_2) - 4\log\frac{1}{\delta'}$$

$$\geq \min\left(1, \frac{5m^2}{k^2}\left(\log\frac{1}{\delta'}\right)^{1-p}\right) \frac{k^p}{(5m)^p}F_p(\mathcal{S}) - 4\log\frac{1}{\delta'}.$$

Setting $\delta' = \delta/(20n)$, we obtain the desired upper bound. ∎

**Lemma 26 (Substream polylog upper bound)** *Let $\mathcal{S} = \langle a_1, a_2, \ldots, a_m \rangle$ be a random order stream on the universe $[n]$. Let $\mathcal{S}' = \langle a_{i_1}, a_{i_2}, \ldots, a_{i_k} \rangle$ be a substream of $k$ fixed distinct integers $K = \{i_1, i_2, \ldots, i_k\}$ with $k \leq m$. Then for fixed $p \in [1, 2], \delta \in (0, 1/2)$, with probability at least $1 - \delta$,*

$$F_p(\mathcal{S}') \leq \left(17\log\frac{20n}{\delta}\right)^p \left(\frac{k}{m}F_p(\mathcal{S}) + 4\log\frac{20n}{\delta}\right).$$

**Proof** Analogous to the proof of Lemma 25, we split the universe items into two parts, $\mathcal{N}_1$ and $\mathcal{N}_2$. For $\mathcal{N}_1$, the corresponding substream is $\mathcal{S}'_1$ and we have that with probability at least $1 - 2n\delta'$,
$$F_p(\mathcal{S}'_1) \leq \frac{3^p k^p}{m^p} F_p(\mathcal{S}_1).$$
For the other part of the universe, $\mathcal{N}_2$, we have that with probability at least $1 - 2n\delta'$, for each $i \in \mathcal{N}_2$,
$$f_i(\mathcal{S}') \leq \frac{k}{m} f_i(\mathcal{S}) + 4\sqrt{\log \frac{1}{\delta}} \sqrt{5 \log \frac{1}{\delta}} \leq 17 \log \frac{1}{\delta'}.$$
Also with probability at least $1 - 2\delta'$,
$$F_1(\mathcal{S}'_2) \leq \frac{k}{m} F_1(\mathcal{S}_2) + 4 \log \frac{1}{\delta'}.$$
Denote $f^* = \max_{i \in \mathcal{N}_2} f_i(\mathcal{S}')$. We obtain,
$$F_p(\mathcal{S}'_2) \leq \sum_{i \in \mathcal{N}_2} f^{*p-1} f_i(\mathcal{S}')$$
$$\leq \left(17 \log \frac{1}{\delta'}\right)^p F_1(\mathcal{S}'_2)$$
$$\leq \left(17 \log \frac{1}{\delta'}\right)^p \left(\frac{k}{m} F_p(\mathcal{S}_2) + 4 \log \frac{1}{\delta'}\right).$$
By setting $\delta' = \frac{\delta}{20n}$, we obtain the desired bound. ∎

**Proof** [of Theorem 27]

Consider a stream of two numbers $a, b \in [n]$. Then there are only two permutations. Namely $\mathcal{S}_1 = \langle a, b \rangle$, $\mathcal{S}_2 = \langle b, a \rangle$. Since $\mathcal{A}$ is correct for at least 60% of the permutations, $\mathcal{A}$ has to be correct on both $\mathcal{S}_1$ and $\mathcal{S}_2$. Therefore, $\mathcal{A}$ can be used to solve the communication problem EQUALITY deterministically, see, e.g., Section 3.5 of [AMS96]. Thus an $\Omega(\log n)$ bits of space lower bound follows from the deterministic communication complexity of the EQUALITy problem. ∎

# B More Proofs on Derandomization

## B.1 Lower Bound

We prove the following lower bound.

**Theorem 27** *For any one-pass deterministic streaming algorithm $\mathcal{A}$, if $\mathcal{A}$ outputs a $(1 \pm 0.1)$-approximation to $F_p(\mathcal{S})$ for at least 60% of all the permutations of stream updates in $\mathcal{S}$, for any $\mathcal{S}$, then $\mathcal{A}$ must use $\Omega(\log n)$ bits of memory.*

**Proof** [Proof of Theorem 27]

Consider a stream of two numbers $a, b \in [n]$. Then there are only two permutations. Namely $\mathcal{S}_1 = \langle a, b \rangle$, $\mathcal{S}_2 = \langle b, a \rangle$. Since $\mathcal{A}$ is correct for at least 60% of the permutations, $\mathcal{A}$ has to be correct on both $\mathcal{S}_1$ and $\mathcal{S}_2$. Therefore, $\mathcal{A}$ can be used to solve the communication problem EQUALITY deterministically, see, e.g., Section 3.5 of [AMS96]. Thus an $\Omega(\log n)$ bits of space lower bound follows from the deterministic communication complexity of the EQUALITY problem. ∎

# C  Other Algorithms

**Algorithm 5:** Estimate the contribution of small frequencies: $\texttt{SmallCont}.(p, n, \epsilon, w_0, L, \{X_i^{(r)}\}, \mathcal{S})$

**Data:**
$L \in \left[\frac{F_p^{1/p}(\mathcal{S})}{C}, F_p^{1/p}(\mathcal{S})\right]$, where $C = \text{poly}(\epsilon^{-1}, \log n)$;

$\mathcal{S} = \langle a_1, a_2, \ldots, a_m \rangle$ is random order stream of length $m$;

$w_0$, the smallest level to be considered;

$\{X_i^r\}$ for $i = 1, 2, \ldots, n$, $r = 1, 2, \ldots, k$ are pairwise independent $p$-Inverse random variables;

**Result:** $R = (r_1, r_2, \ldots, r_k)$, where each $r_i \in \mathbb{R}$;

1 **Initialize:**
2 $\quad Z_0^{(r)} \leftarrow \left\{i \in [n] \times [k] : X_i^{(r)} \geq CL\right\}$; $Z_w^{(r)} \leftarrow \left\{i \in [n] \times [k] : \frac{CL}{2^{(l-1)/p}} < X_i^{(r)} \leq \frac{CL}{2^{w/p}}\right\}$;
3 $\quad \mathcal{S}_w^{(r)} \leftarrow S|_{Z_w^{(r)}}$ the substream induced by $Z_w^{(r)}$, and $\mathcal{S}_w^{m_1:m_2(r)} \leftarrow S^{m_1:m_2}|_{Z_w^{(r)}}$;
4 $\quad CS_w \leftarrow \texttt{CountSketch}\left(\epsilon, \frac{\epsilon^2}{w_0}\right)$ /*Modified countsketch instance such that each counter is restricted to use $\mathcal{O}\left(\log \frac{\log n}{\epsilon}\right)$ bits.*/;

5 Run in parallel:
6 **for** $w \in [w_0]$ **do**
7 $\quad$ Hash the IDs of $Z_w$ to the universe $[\text{poly}(\log n, \epsilon^{-1})]$;
8 $\quad CS_w.\text{input}(\mathcal{S}_w)$;
9 $\quad$ Run in parallel:
10 $\quad$ **for** $r \in [k]$ **do**
11 $\quad\quad B \leftarrow \texttt{BPTree}.\text{initialize}\left(\frac{1}{w_0}, \frac{0.01}{w_0}\right)$; $m_0 \leftarrow 0$;
12 $\quad\quad$ **while** $\textit{sizeof}(\textit{memory}(B)) \leq \Theta\left(w_0 \log w_0 \left(\log \log n + \log \frac{1}{\epsilon}\right)\right)$ **do**
13 $\quad\quad\quad B.\text{update}(\text{stream updates of } \mathcal{S}_w^{(r)})$;
14 $\quad\quad\quad m_0 \leftarrow m_0 + 1$;
15 $\quad\quad H_w \leftarrow B.\text{return}()$;
16 $\quad\quad$ **for** $i \in H_w$ **do**
17 $\quad\quad\quad f_i \leftarrow \frac{m}{m_0} f_i\left(\mathcal{S}^{m_0:2m_0}\right)$; /*$m$ is calculated at the end of the stream.*/;

18 $\quad$ At the end of the stream:
19 $\quad$ **for** $i \in H_w$ **do**
20 $\quad\quad f_i' \leftarrow CS_w(i)$;
21 $\quad\quad$ **if** $f_i \neq \infty$ **then**
22 $\quad\quad\quad f_i \leftarrow f_i'$;

23 **return** $\bigcup_{w=0}^{w_0} H_w$;

**Algorithm 6:** Estimate the frequency of an element: `QueryFrequency`$(i^*, \widehat{m}, \epsilon, \delta, \mathcal{S})$

**Data:**
    $\mathcal{S} = \langle a_1, a_2, \ldots, a_m \rangle$ is a random order stream;
    $\epsilon, \delta \in (0, 1/2)$;
    $i^* \in [n]$, the frequency to be queried;
    $\widehat{m}$ : the length of the stream used to query the frequency of item $i^*$;

**Result:** $\widehat{f} \in \mathbb{R}$;

1 **Query:**
2     **if** $|\mathcal{S}| < \widehat{m}$ **then**
3         **return** 0;
4     **else**
5         **return** $\frac{f_{i^*}(\mathcal{S}^{1:\widehat{m}})}{\widehat{m}}$;

---

**Algorithm 7:** $F_p$-algorithms for small streams: `SmallApprox`$(p, n, \mathcal{S})$

**Data:**
    $\mathcal{S} = \langle a_1, a_2, \ldots, a_m \rangle$ is a random order stream of length $m$;

**Result:** $F$: $F \in ((1-\epsilon)F_p(\mathcal{S}), (1+\epsilon)F_p(\mathcal{S}))$ is a $(1 \pm \epsilon)$-approximation to $F_p(\mathcal{S})$;

1 Run an optimal turnstile streaming algorithm $\mathcal{A}$ for $F_p$, with the following modification: **if** *at any point the counter exceeds* poly $\left(\frac{1}{\epsilon}\right)$ poly $\log n$ **then**
2     d
3 e **return** `Fail`. **else**
4     **return** whatever the algorithm $\mathcal{A}$ returns.

---

**Algorithm 8:** $F_p$-Heavy Hitter Algorithm: $\text{HHR}(p, c_0, \widehat{m}, F, \mathcal{S})$

---
**Data:** ;
    $0 < c_0 \leq 1$ is a constant;
    $p \in (0, 2)$ a real number;
    $\widehat{m} \geq \frac{m}{C}$, for some constant parameter $C > 1$;
    $F \in \left[\frac{F_p(\mathcal{S})}{C_1}, F_p(\mathcal{S})\right]$, for some parameter $C_1 = \text{poly}(\epsilon^{-1}, \log n)$;
    $\mathcal{S} = \langle a_1, a_2, \ldots, a_m \rangle$ is a random order stream of length $m$;
**Result:** $\mathcal{H} \subset [n]$: a set of $(c_0, F_p)$-heavy hitters.

**1** **Initialize:**
**2**     $m_1 \leftarrow \frac{\widehat{m}^2}{F^{2/p}}$;
**3**     $t \leftarrow c_1 \log n$; /*for some sufficiently large constant $c_1 > 0$*/;
**4**     $M \leftarrow \text{MG}(c_2)$; /*a Misra-Gris algorithm instance, for small enough constant $c_2 \leq c_0$*/;
**5** **if** $m_1 \leq (\log n)^{c_3}$ **then**
**6**     $m_1 \leftarrow (\log n)^{c_3}$; /* $c_3 > 0$ is a large enough constant*/;
**7** **while** *not at end of the stream* **do**
**8**     Let $\mathcal{S}_1, \mathcal{S}_2, \ldots, \mathcal{S}_t$ be consecutive stream updates each of length $m_1$;
**9**     **for** $j = 1$ *to* $t$ **do**
**10**         $H_j \leftarrow \text{BPTree}(\mathcal{S}_j, c_2, 0.01)$;
**11**         /*Use the BPTree to detect constant heavy hitters, errs with probability at most $0.01$*/;
**12**         $M.\text{update}(H_j)$; /*Input all elements in $H_j$ to MG*/;
**13**     **return** $M.\text{outputs}()$.
**14** **return** *Fail*.

---